\documentclass{article}
\usepackage[utf8]{inputenc}
\usepackage{blindtext}
\usepackage{hyperref}
\usepackage[a4paper]{geometry}
\usepackage{amsmath}
\usepackage{amssymb}
\usepackage{mathabx}
\usepackage{graphicx}
\graphicspath{{figures/}} %Setting the graphicspath
\usepackage{subfig}
\usepackage{enumitem}
\usepackage{csquotes}
\usepackage[
    url=false,
    doi=false,
    isbn=false,%    eprint=false,
    style=ieee,  
    backend=biber
  ]{biblatex} %Imports biblatex package
\AtEveryBibitem{%
  \clearlist{language}%
}
\usepackage{amsthm}
\usepackage{amsfonts}
\usepackage{amssymb}
\usepackage{mathrsfs}
\usepackage{mathtools}
\usepackage{xcolor,soul}
\usepackage{tikz}
\usepackage{pifont}
\usepackage{relsize}

\usetikzlibrary{decorations.pathreplacing}

\newcommand{\coeff}{a}
\newcommand{\eps} {\varepsilon}
\newcommand{\eul}{\mathrm{e}}
\newcommand{\imu}{\mathrm{i}}
\newcommand{\bigo}[1]{\mathcal{O}\left(#1\right)}
\newcommand{\diff}[2]{\frac{\mathrm{d} #1}{\mathrm{d} #2}}
\newcommand{\diffd}{\mathrm{d}}

\newcommand{\myfloor}[1]{\left \lfloor #1 \right \rfloor }

\newcommand{\abbeq}{Eq. }
\newcommand{\borel}[1]{\mathcal{B}\left[ #1\right]}
\newcommand{\bpade}[2]{\text{BP}_{#2}\left[ #1\right]}

\title{The late to early time behaviour of an expanding plasma: \\ 
 hydrodynamisation from exponential asymptotics}
\author{In\^{e}s Aniceto$^{1}$, Daniel Hasenbichler$^1$, Adri Olde Daalhuis$^2$}
 \date{%
     $^1$School of Mathematical Sciences, University of Southampton, Southampton, UK\\[2ex]%
    $^2$School of Mathematics and Maxwell Institute for Mathematical Sciences, \\[0.7ex]
    The University of Edinburgh, Edinburgh, UK
}         

\begin{document}

\maketitle

\abstract{
We use exponential asymptotics to match the late time temperature evolution of an
expanding conformally invariant fluid to its early
time behaviour. We show that the rich divergent transseries
asymptotics at late times can be used to interpolate
between the two regimes with exponential accuracy 
using the 
well-established methods of hyperasymptotics,
Borel resummation and transasymptotics. 
This approach is generic and can be applied 
to any interpolation problem involving a local asymptotic transseries expansion as well as knowledge of the solution in a second region away 
from the expansion point.
Moreover, we
present \textit{global analytical properties}
of the solutions 
such as analytic approximations to 
the locations
of the square-root branch points, exemplifying how the summed transseries contains within itself information
about the
observable in regions with different asymptotics.
}

%%%%%%%%%%%%%%%%%%%%%%%%%%%%
\section{Introduction}
\label{sec:introduction}
%%%%%%%%%%%%%%%%%%%%%%%%%%%%

Viscous relativistic hydrodynamics is a long-wavelength effective 
theory which has been traditionally thought to be 
valid only near local thermal equilibrium. Surprisingly,
hydrodynamic models can be successfully applied to 
certain physical systems which are far from equilibrium,
such as an expanding quark-gluon plasma created from
heavy-ion collisions at relativistic energies 
\cite{
Shen:2020mgh,
gale2013hydrodynamic,
Romatschke:2009im,
Heinz:2013th}.
In those
cases the hydrodynamic model contains within itself
emergent, non-hydrodynamic degrees of freedom which are
non-perturbative in nature and decay exponentially 
in time toward a hydrodynamic attractor \cite{heller2015hydrodynamics}. This process
is known as hydrodynamisation (see \textit{e.g.} \cite{Florkowski:2017olj}).  These non-hydrodynamic modes play a major role during the early times of the expanding plasma, and are quite sensitive to the different initial conditions. At the hydrodynamisation time, the system is still far from equilibrium and its pressure is quite anisotropic, but nevertheless the different initial solutions all become exponentially close to each other, and the evolution of the system towards equilibrium is effectively described by viscous hydrodynamics, via the same power series expansion in small gradients valid at late times.

From the point of view of asymptotics, however, such behaviour is expected. Mathematically,
the late-time attractor is described by a 
divergent, asymptotic perturbative
series, whose resurgent properties encode all
the information about the 
non-perturbative modes. The information about the initial conditions is instead uniquely encoded in a set of parameters determining the strength of the (exponentially small) non-hydrodynamic modes.\footnote{The
number of parameters will be in one-to-one 
correspondence to the elementary 
non-hydrodynamic modes $\sigma_i \, \mathrm{e}^{-A_i w}$.} The full description of the system can then be achieved via a so-called resurgent transseries.

Thus, when solving
for the time-evolution of an observable
(generally determined by some ODE/PDE within
the hydrodynamic model), the initial-conditions (constrained by the physics) can dramatically change the behaviour at early-times by fixing the strength of the non-hydrodynamic modes dominating this early-time regime, 
while at late times
the non-hydrodynamic modes are exponentially suppressed
and thus negligible in terms
of their numerical magnitude, washing away the information on the initial conditions, with only the hydrodynamic power-law decay towards the attractor remaining.

Having access to the behaviour of our system at an initial time, as well as a description of its late-time asymptotic behaviour, one is naturally left with a few questions: \textit{how can we match the late-time behaviour
to any given initial condition? Beyond a purely numerical analysis of the observable, how can we use this matching
to describe the system at all times? Can we hope to describe the analytic behaviour of our observable?}

The factorial divergent nature of the late-time expansions and their resurgent properties provides a path to answer these questions. Unlike many of their convergent counterparts, it is well known that these asymptotic expansions converge to their expected results quite quickly -- in fact keeping just a few terms provides a very precise approximation, which can be extended far beyond the original expansion point, in our case large time 
(see \textit{e.g.} \cite{bender1999advanced}). 
Moreover, there are well established asymptotic summation methods, based on the underlying resurgent properties (see \textit{e.g.} \cite{Aniceto:2018bis} and references therein),  which provide such an approximation with exponential accuracy, thus effectively distinguishing between the exponentially close solutions at late-times 
\cite{caliceti2007useful,
berry1990hyperasymptotics,
olde1995hyperasymptotic}. As we will see, some of these methods even allow us to study global analytic properties of the asymptotic observable in its domain of interest, such as existence of poles or branch points   
\cite{costin1999correlation,costin2001formation,costin2015tronquee,
upcoming_paper_painleve}

Naturally one should start with how to interpolate our late-time solution with a given initial condition. Unlike previous work discussing this matching in the context of relativistic hydrodynamics \cite{
behtash2019dynamical,
behtash2021transasymptotics}, where the interpolation was done using a
numerical least-square fit, our approach will involve various resummation methods based on the resurgent properties of the late-time solution. We will show that these resummation methods
can be used to calculate the residual parameter labelling the exponentially close 
late-time solutions, being also highly effective at 
computing the solutions with exponential accuracy. Hence they
are excellent approximations
for most times outside of some region at very early times,
where all orders of the non-hydrodynamic exponential modes are of notable size and drive the behaviour of the system.

%%%%%%%%%%%%%%%%%%%%%%
\subsection*{A simple model of hydrodynamics}
%%%%%%%%%%%%%%%%%%%%%%

We will solve the interpolation problem
between late and early times for an ODE
describing the evolution of the
effective 
temperature \footnote{The system is not at thermal equilibrium, 
hence 
strictly speaking there is no temperature.
The effective temperature is defined
as the temperature of a system at thermal
equilibrium with the same energy density.
In the rest of this paper we will nonetheless
use the term 'temperature' to refer to $T$.}
of a conformal fluid 
in $d=4$ dimensions undergoing a boost-invariant expansion. The
model can be regarded 
as a toy model for the expansion
of a strongly-coupled
Quark-Gluon-Plasma created after a 
collision of two heavy ions beams. We assume
rotational and
translational invariance transverse to
the collision axis. Further, we
assume boost invariance with respect
to boosts along the collision axis
(Bjorken flow, see \cite{bjorken1983highly}),
which is a reasonable approximation 
at high energies in the central rapidity
region.
Hence all observables in our system only
depend on the proper time $\tau$
of some inertial observer
and we may write $T = T(\tau)$
for the temperature. The energy momentum tensor
of our system is given by
\begin{equation}
    \label{eq:energy_momentum_tensor}
    T^{\mu \nu} = \mathcal{E} \, u^\mu u^\nu
    + (g^{\mu \nu} + u^\mu u^\nu) p(\mathcal{E})
    + \Pi^{\mu \nu},
\end{equation}
where $g^{\mu \nu} = \text{diag}(-1,1,1,1)^{\mu \nu}$
is the flat Minkowski metric,
$\mathcal{E}$ is the energy density
in the rest frame of the fluid, $p(\mathcal{E}) = \mathcal{E}/3$
is the pressure of
a perfect conformal fluid,
the vector
$u^\mu$ is the four-velocity of the fluid,
and $\Pi^{\mu \nu}$ is the shear-viscosity
tensor. Conservation of energy and
conformal symmetry imply (the symbol
$T$ in the third equation stands for temperature):
\begin{equation}
    \label{eq:energy_conservation_trace_constraint}
    \nabla_\mu T^{\mu \nu} = 0; \quad \quad 
    {T^\mu}_\mu =0;
    \quad \quad \mathcal{E} \sim T^4 \,.
\end{equation}
The most straightforward approach
towards
solving 
\eqref{eq:energy_conservation_trace_constraint}
for the temperature $T(\tau)$
is to expand the shear-stress tensor 
$\Pi^{\mu \nu}$ from 
\eqref{eq:energy_momentum_tensor}
by summing all the allowed derivative terms
up to a given order.
However, 
the equations one obtains are not hyperbolic,
and hence the model is acausal. An alternative 
way of dealing with  
\eqref{eq:energy_conservation_trace_constraint}
is to upgrade the shear-stress tensor
$\Pi^{\mu \nu}$ to an independent
field satisfying a relaxation-type
differential equation. This
approach, called 
Müller-Israel-Stewart (MIS) theory,
\cite{muller1967paradoxon,israel1979transient,ruggeri1986relativistic,geroch1991causal}
results in a causal model
and is the one we 
will use in this work 
(see \cite{heller2015hydrodynamics,
bacsar2015hydrodynamics,
aniceto_resurgence_2016}).
Instead of using the 
variables $(\tau,T)$, where $T$ stands for 
the temperature,
it is more convenient to work with the
variables $(w,f)$
defined by
\footnote{
Our definition of $f(w)$ in (\ref{eq:var_transformation_w_f})
differs from the convention in \cite{heller2015hydrodynamics}
by $f_\text{ours} = \frac{3}{2} f_\text{theirs} $.
We nonetheless chose this normalisation because it leads to simpler
equations.
}:
\begin{equation}
	\label{eq:var_transformation_w_f}
w = T \tau; \quad \quad \quad  f = \frac{3 \tau}{2 w}\diff{w}{\tau}.
\end{equation}
The variable $w$ measures proper time in units of inverse temperature, and
the quantity $f$ is closely related to the pressure
anisotropy
\footnote{The pressure anisotropy $\mathcal{A}$
is related to $f$ by
 $\mathcal{A} = 8 \left( f -1 \right)$
 \cite{Heller:2011ju}.
}.
The differential equation
describing the evolution of $f(w)$ in MIS theory 
is
\begin{equation}
	\label{eq:ode_mis_model_f_of_w}
	w f(w) f^\prime(w) +4 f(w)^2+ f(w) \left( -8+A w \right)+\left(4-\beta-A w \right) =0 .
\end{equation}
The parameters $A$ and $\beta$ depend on
phenomenological
constants. If the microscopic theory behind a physical system is
known, it can be used to derive these parameters. Our analysis can be
performed in exactly 
the same way for any values of $A$ and $\beta$. 
However, in 
this work we will only work with
the following values for $A$ and $\beta$
\footnote{Regarding our choice of parameters:
The three phenomenological parameters defining the
second-order transport coefficients which are relevant for the MIS dynamics
are
$C_{\tau \Pi}$, $C_{\lambda_1}$, and $C_\eta$. 
Assuming the microscopic theory is $\mathcal{N}=4$ SYM  these parameters have been derived using holography and are given by
 \cite{bhattacharyya2008nonlinear,
baier2008relativistic}:
\begin{equation*}
	\label{eq:n=4_sym_transport_coefficients}
C_{\tau \Pi}^{\text{(SYM)}} = \frac{2-\log 2}{2\pi}; 
\quad C_{\lambda_1}^{\text{(SYM)}} = \frac{1}{2\pi}; \quad
C_\eta^{\text{(SYM)}} = \frac{1}{4\pi};
\end{equation*}
The ODE (\ref{eq:ode_mis_model_f_of_w}) is obtained by setting
$C_{\lambda_1} = 0$,  and identifying $A= \frac{3}{2 C_{\tau \Pi}}$ and
$\beta = \frac{C_\eta}{C_{\tau \Pi}}$. We followed \cite{heller2015hydrodynamics} and chose $C_{\lambda_1} = 0$
because this special case leads to a very interesting mathematical 
structure. The other two phenomenological parameters are chosen
as $C_{\tau \Pi} =C_{\tau \Pi}^{\text{(SYM)}}$, $C_{\eta} =C_{\eta}^{\text{(SYM)}}$,
leading to (\ref{eq:values_a_and_beta}). Note that Eq.
\eqref{eq:ode_mis_model_f_of_w}
is only correct in the case $C_{\lambda_1} = 0$.
}:

 \begin{equation}
	\label{eq:values_a_and_beta}
	A = \frac{3 \pi }{2-\log 2}; \quad \quad \quad \beta = \frac{1}{2 (2-\log 2)};
\end{equation}

\begin{figure}[ht!]
\centering
\includegraphics[width=0.4\linewidth]{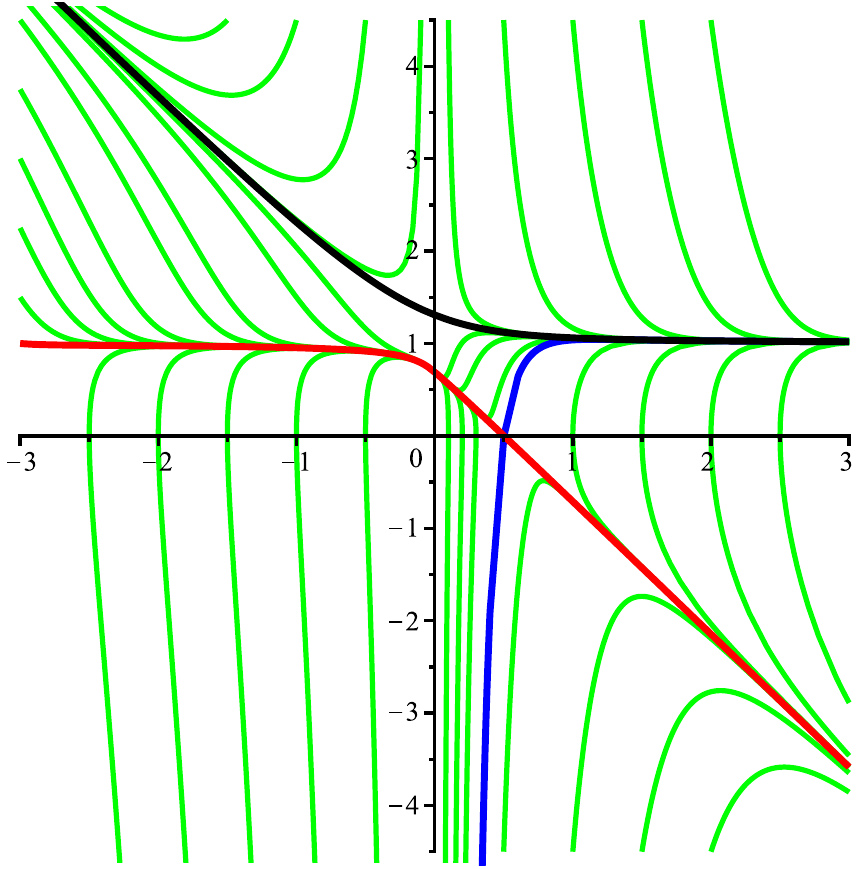}
\qquad\includegraphics[width=0.4\linewidth]{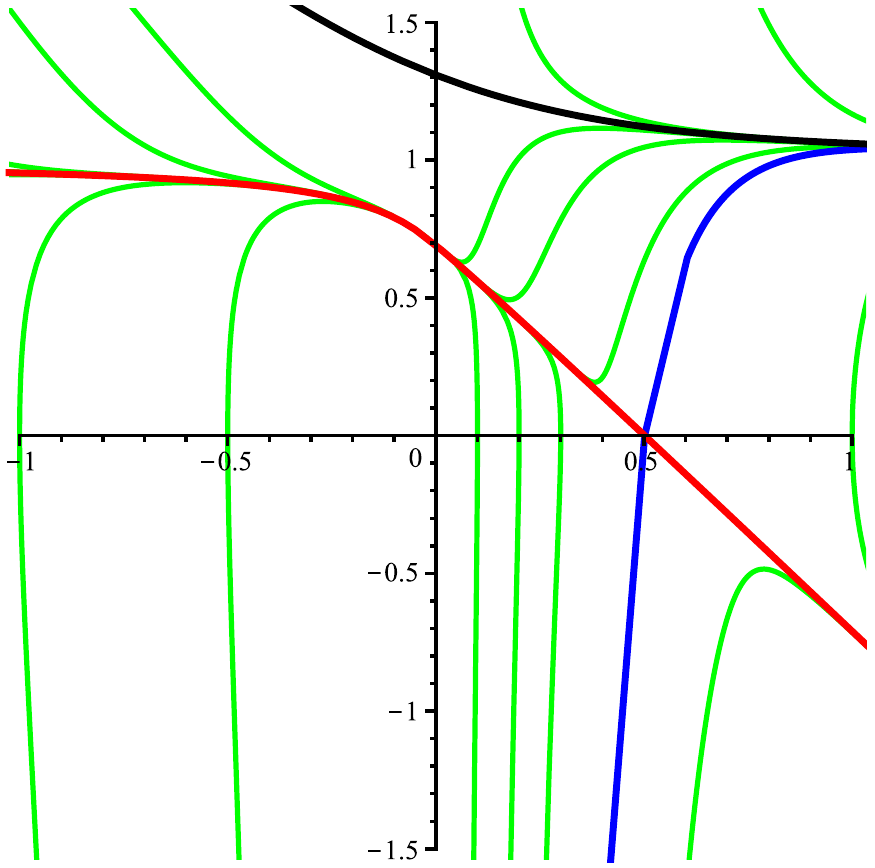}
\caption{The real graph $(w,f(w))$ plane. The figure on the right is a zoom-in around the origin of the figure on the left. 
The red and blue curves are the only two solutions with a regular zero. 
The red curve representing $f_-(w)$ and black  curve representing $f_+(w)$ are the only two solutions with finite values
at the origin.}
\label{fig:regularzero}
\end{figure}

Let us consider the solutions of \abbeq (\ref{eq:ode_mis_model_f_of_w}).
In the solutions plot Fig. \ref{fig:regularzero}
the real solution along the real axis are displayed. 
There are two distinct solutions, represented by the 
red and black curves in  Fig. \ref{fig:regularzero},
which are finite at
the origin. We will call the solution represented by
the black curve $f_+$,
and the one represented by the red curve $f_-$. 
The functions $f_+$ and $f_-$ are special solutions because they are attractors 
at $w=+\infty$, and $f_+/f_-$ is the attractor/repellor at $w=-\infty$, respectively.
This means that all other solutions, represented by green curves in Fig.
\ref{fig:regularzero}, get exponentially close to either $f_+(w)$ or $f_-(w)$
as $w \to +\infty$. An important feature
of the solutions is the presence of square root
branch points, whose locations we shall denote
by $w_\text{s}$. It can be shown that \abbeq 
(\ref{eq:ode_mis_model_f_of_w}) admits solutions
of square root type, and admits the following analytic
expansion in the variable $(w-w_\text{s})^{1/2}$:
\begin{equation}
	\label{eq:f_of_w_branch_point_sols}
	 f(w)=\sum_{n=1}^\infty h_n(w_\text{s})\left(w-w_\text{s}\right)^{n/2},
	 \qquad\textrm{with}~~ h_1^2(w_s)=\frac{2Aw_s+2\beta-8}{w_s},~~
	 h_2(w_s)=\frac{16-2Aw_s}{3w_s},~~\cdots.
\end{equation}

The locations $w_\text{s}$ of the branch points depend on the
initial conditions we impose on $f(w)$. The presence of 
these square root branch points implies that the natural domain 
of $f(w)$ is a non-trivial Riemann surface. 
Note that the summation in \eqref{eq:f_of_w_branch_point_sols} starts at $n=1$,
hence, those solutions are zero at the branch point. It is easy to check
that the only possible regular zero for solutions of \eqref{eq:ode_mis_model_f_of_w}
is at $w=(4-\beta)/A\approx 0.5$, the point of intersection of red and blue
curve in Fig. \ref{fig:regularzero}.
Let us now
analyse how the real solutions  are related to
each other by considering their expansions
around the origin and around infinity.
\subsection*{Solutions around the origin}
Around $w=0$ we
have the following convergent expansions:
 \begin{enumerate}[label=(\alph*)]
\item \begin{equation}
\label{eq:f_plus_minus_leading_order}
f_\pm(w)= (1\pm\sqrt{\beta}/2)+\bigo{w} , \quad w \to 0,
\end{equation}
\item \begin{equation}
\label{eq:f_c_leading_order}
f_C(w) =C w^{-4} +2+ \bigo{w},  \quad w \to 0.
\end{equation}
\end{enumerate}
There is a relationship between
$f_+(w)$ and $f_C(w)$. In the solutions 
plot Fig. \ref{fig:regularzero}
the green curves above the graph 
of $f_+(w)$ (in black) correspond
to $f_C(w)$ for $C>0$. We can see 
that as $C$ becomes smaller, 
the green curves in Fig. \ref{fig:regularzero} get closer
to $f_+(w)$. In the limit $C\to 0^+$, $f_C(w)$ converges
pointwise to $f_+(w)$ for $w \neq0$. Hence $f_+(w)$
can be understood as the $C \to 0^+$ limit of $f_C(w)$,
in which the divergence at the origin disappears.
For $0>C>C_{\text{split}}\approx -0.0874$, $f_C(w)$
has a square root branch point on the positive 
real axis. At
$C=C_\text{split}$ this square root 
singularity splits into two singularities, one
above and one below the real axis. 
The corresponding function
$f_{C_\text{split}}(w)$ is represented 
by the blue curve in Fig. \ref{fig:regularzero}. 
For $C<C_\text{split}$, the function
$f_C(w)$ has no square root branch points
along the real axis and admits a real solution
for all $w>0$.
These solutions are 
represented by the green curves in the bottom right corner of
Fig. \ref{fig:regularzero}.

\subsection*{Solutions around infinity}
Around $w =\infty$ we also have two distinct
transseries expansions,
depending on whether the solutions
converge or diverge at infinity.

 \begin{enumerate}[label=(\alph*)]
\item \textit{Solutions of finite limit:} the
solutions  which converge to the hydrodynamic 
attractor $f_+(w)$ as $w\to +\infty$
can be described with the following transseries
expansion \cite{heller2015hydrodynamics}:
\begin{equation}
	\label{eq:transseries_f}\mathcal{F}(w,\sigma) =
	 \sum_{n=0}^{\infty} \sigma^n w^{\beta n} 
	 \eul^{-n A w} \Phi^{(n)}(w).
	\end{equation}

\noindent
The transseries $\mathcal{F}$ describes a one-parameter family
of solutions converging to the finite value $\mathcal{F} \to 1$ as
$w \to +\infty$. Hence all the green curves in Fig. \ref{fig:regularzero}
which approach the black curve $f_+(w)$ have a transseries parameter
$\sigma$ assigned to them. 
The parameter $\sigma$ is undetermined by the equation and has to 
be matched with the early-time behaviour of $f(w)$ around $w=0$,
around which point we know all the solutions as convergent series
expansions \eqref{eq:f_plus_minus_leading_order} and 
\eqref{eq:f_c_leading_order}. 
We will determine later in the paper that the $\sigma$ corresponding
to the black/blue curve in Fig. \ref{fig:regularzero} 
are approximately $\sigma_+ =-0.3493 + 0.0027 \imu$
and $\sigma_{\rm blue} =-14.4111 + 0.0027 \imu$, respectively.
The particular form of \eqref{eq:transseries_f}
implies that we know the amplitudes of all the non-perturbative modes
once we know the transseries parameter $\sigma$.
The expression $\Phi^{(n)}(w)$ stands for the divergent, asymptotic series
of the $n$-th non-perturbative sector or non-hydrodynamic mode: 
\begin{equation}
	\label{eq:phi_n_expansion}
	\Phi^{(n)}(w) = \sum_{k=0}^{\infty}  \coeff^{(n)}_k w^{-k}.
\end{equation}

\noindent
The coefficients $\coeff_k^{(n)}$ above can be determined from recurrence equations obtained by using the ansatz (\ref{eq:transseries_f})
into the MIS ODE (\ref{eq:ode_mis_model_f_of_w}),
and matching equal powers of $\sigma$ 
(see Appendix \ref{ap:recurrence_relations}). We use
the convention $\coeff^{(1)}_0=3/2$. \footnote{With
this convention our Stokes constant and
transseries-parameter normalisation is the same as in
\cite{heller2015hydrodynamics,bacsar2015hydrodynamics,
aniceto_resurgence_2016}, and choosing a
different value for $\coeff^{(1)}_0$ corresponds
to a rescaling of $\sigma$.} 
 The hydrodynamic series $\Phi^{(0)}(w) = 1 + \frac{\beta}{A}w^{-1} + \bigo{w^{-2}}$ describes the
perturbative sector and defines the attractor. Due to the factor $\eul^{-n A w}$ 
multiplying the non-hydrodynamic series $\Phi^{(n)}(w), \, n\geq 1$,
the convergence of the solutions
to the attractor is 
exponentially fast.
 
 \item \textit{Growing solutions:}
the solutions which are linearly growing to leading
 order and asymptotically
 approximate $f_-(w)$ as $w \to +\infty$ admit the following transseries
 expansion
 \footnote{Note that the transseries $\Psi$ 
 from \abbeq \eqref{eq:p4_family_solutions} is constructed 
 from the basis monomials
$w^{-1}$ and $\log w$, 
whereas $\mathcal{F}$ is constructed from the
basis monomials $w^{-1}$ and $\eul^{-A w}$.
}
 \begin{equation}
	\label{eq:p4_family_solutions}
	\Psi(p_4,w) = \sum_{k=-1}^{3}  p_k \,w^{-k}
	+ \sum_{k=4}^{9} w^{-k}\left( p_k +q_k \log w\right)
	+\sum_{k=10}^{14} w^{-k} \left( p_k +q_k \log w+r_k \log^2 w\right)
	+\dots ,
\end{equation}
with $p_{-1}=-A /5$.
The first four coefficients, $p_n, \, -1\leq n \leq 3$, are uniquely determined
by the MIS equation \eqref{eq:ode_mis_model_f_of_w} alone. 
The coefficient $p_4$ is undetermined by  \eqref{eq:ode_mis_model_f_of_w},
and all other coefficients generically depend non-linearly
on the coefficient $p_4$.
Hence the transseries $\Psi$ from \eqref{eq:p4_family_solutions} represents
a one-parameter
family of solutions. 
The red curve in Fig. \ref{fig:regularzero}, that is $f_-$, corresponds to $p_4=-0.3474942558$.
It should be obvious from Fig. \ref{fig:regularzero} that as $w\to-\infty$ the regular
solutions have a transseries expansion of the form \eqref{eq:p4_family_solutions}.
\end{enumerate}
The exponential transseries (\ref{eq:transseries_f})
can be regarded as
an expansion with a two-scale structure, the
perturbative variable $w^{-1}$, as well as an
exponential variable
\begin{equation}
    \label{eq:def_tau}
    \tau \equiv \sigma\, w^{\beta}
    \mathrm{e}^{-Aw}.
\end{equation}
Notice the form of the  
transseries 
(\ref{eq:transseries_f}):
the outer sum is performed
over powers of $\tau$, with
coefficients $\Phi^{(n)}(w)$ 
depending on the variable
$w^{-1}$. 
In this work we will present different summation approaches of the asymptotic functions $\Phi^{(n)}(w)$. We will also explore an alternative way of summing (\ref{eq:transseries_f})
called transasymptotic  summation, 
in which the order
of summation is reversed \cite{costin2001formation}:
the coefficients in the 
$w^{-1}$-expansion are then functions of
$\tau$ (defined by a convergent Taylor series in $\tau$).
Thus the
divergence of the transseries is not caused by
the large-order behaviour of the exponential scales, but
instead by the divergent asymptotic expansions 
at each order of the
non-perturbative exponential.
Although
the transseries \eqref{eq:transseries_f}
was an expansion
around $w=+\infty$, the transasymptotic approach
allows us to access different regimes where
$\tau$ is no longer small.

%%%%%%%%%%%%%%
\subsection*{Outline}
%%%%%%%%%%%%%%%

In Section 2  we perform
the exponentially accurate
($\text{error} \sim \eul^{-2|A w|}$ ) interpolation
between late and early times with
two different asymptotic methods:
hyperasymptotics and Borel resummation.
In particular, we show how to compute
the transseries parameter
$\sigma$ from \eqref{eq:transseries_f} with
accuracy $~ \eul^{- |A w|}$ from the 1-parameter family of of solutions at initial time. We
explain how the matching
function $\sigma(C)$ can be used 
to illustrate the
convergence of the initial solutions
$f_C(w) \to f_+(w)$ as $C \to 0$
(see Fig. \ref{fig:sigma_plot}).
This matching is performed at a chosen matching point, and the analytic continuation 
$f_\text{ac}(w)$ from the
origin to the matching point 
is performed numerically
using the Taylor series method.
In Section 3 we introduce the 
transasymptotic summation and derive 
an asymptotic expansion 
for $\sigma$ in closed-form 
at the matching point.
Although the accuracy is worse
with respect to
hyperasymptotics and
Borel resummation, it allows us to
obtain analytic results which are
useful far beyond the interpolation
problem between late and
early times and which
can be employed to deduce
\textit{global} properties of the
solutions.  For example,
one can derive an asymptotic formula
for the location
of square-root branch points, and
explain the
differing asymptotic
expansions in two different regions
of our complex domain $(w\rightarrow \pm\infty$) as a direct
consequence of the change in 
sign of the exponents $\log(\tau^n) \sim 
- n A \, w$ of
our non-perturbative exponentials.

%%%%%%%%%%%%%%%%%%%%%%%%%%%%%%
\section{Interpolation between late-times and early-times}
	\label{sec:late_and_early_times}
%%%%%%%%%%%%%%%%%%%%%%%%%%%%%%%

In the previous section we described
the behaviour
of the solutions to the MIS equation
\eqref{eq:ode_mis_model_f_of_w}
both for early- and late-times.
We found that there
exists a one-parameter
family of solutions with a finite
limit at infinity. These
solutions 
converge exponentially fast
to a hydrodynamic
attractor described by a perturbative series.
We saw that this series could be upgraded
to the transseries 
\eqref{eq:transseries_f}
by including decaying exponential terms
at large times. The transseries
parameter $\sigma$ from \eqref{eq:transseries_f}
was identified as 
a proxy for the amplitudes of
the non-perturbative exponential modes.
In the early-time regime near the origin,
we found another representation of 
said family of
solutions \eqref{eq:f_c_leading_order},
labelled by the leading-order
coefficient $C$ of their Laurent expansion around
the origin. 
Linking
the magnitude of the
non-perturbative modes of the late-time
asymptotic transseries to the early-time
behaviour can be very useful, and has
previously been done
by numerical fitting
\cite{behtash2021transasymptotics,
Behtash:2019txb}. 
However, the fitting
approach does not exploit the vast possibilities arising
from the rich late time asymptotics of the solutions. 
In particular, the
difficulty with the fitting method
lies in the exponential proximity of
any two distinct solutions at late times, and hence
a significant deviation from the desired solution is
weakly penalised at late times, while at early times
the function is not accurately 
captured by the fit model due to the finite truncation
of the transseries \eqref{eq:transseries_f}.
Fortunately, given that our solution at late times 
is divergent asymptotic, we have a range of 
tools at our disposal to do the matching,
whose exponential 
accuracy provides a way to differentiate 
the behaviour of the different solutions.
The matching  between late and early times will be achieved in 
three main steps:

 \begin{enumerate}[label=(\roman*)]
\item we will sum the factorial divergent expansion at late times, using exponentially accurate methods, keeping not only the perturbative series but also a non-perturbative, exponentially small part (effectively keeping the exponential accuracy). We will then evaluate this sum at a finite but large enough time $w_0$.
\item we will analytically continue the solution at the origin to the same value $w_0$.
\item the two approximations we will find depend on their respective parameters ($C$ representing early times
and $\sigma$ late times) and their relation can be obtained via direct comparison.
\end{enumerate}
After having found the transseries parameter for a given
solution at early times, we can use the asymptotic 
summation methods
to find exponentially accurate interpolations
in the regime between early-times and infinity.

%%%%%%%%%%%%%%%%%%%%%%%%%%%
\subsection{Hyperasymptotic summation}
%%%%%%%%%%%%%%%%%%%%%%%%%%%

Hyperasymptotics is a resummation method which
exploits the asymptotic properties of the transseries
to approximate the value of a function by 
truncated sums 
\cite{berry1990hyperasymptotics,
berry1991hyperasymptotics,
olde1995hyperasymptotic,
olde1995hyperasymptotic2}. In computing our approximation for  the solution $f(w_0)$ 
to the MIS ODE \eqref{eq:ode_mis_model_f_of_w} 
at a finite 'matching time' $w_0$ from the late time solution (transseries), we will keep terms up to linear order in the transseries parameter $\sigma$
from \eqref{eq:transseries_f}. 
This corresponds to calculating level-one hyperasymptotics, 
for which we need to 
compute terms of the transseries sectors
$\Phi^{(0)}(w)$ and $\Phi^{(1)}(w)$ from the
transasymptotic summation 
\eqref{eq:transseries_f} (see Appendix \ref{ap:recurrence_relations} for the
computation).
The optimal number of terms $N_\text{Hyp}$
at which the series expansions arising in
level-one hyperasymptotics must be truncated
is a function of the resummation
point $w$  at which we 
 wish to resum the transseries 
 \cite{olde2005hyperasymptotics}:
\begin{equation}
	\label{eq:n_hyp_formula}
N_\text{Hyp}(w) = 2 \myfloor{\big| A w\big|};
\end{equation}

\noindent
where $\myfloor{\cdots}$ is the usual floor function. Thus we need to 
compute the terms of the power series
$\Phi^{(0)}(w)$ and $\Phi^{(1)}(w)$ to sufficiently
high order (we used a maximum of $100$ terms
\footnote{Using 200 terms allows us to use the hyperasymptotic
approximation with optimal precision up to $w = 7$.
}for all our approximations).

The level-one hyperasymptotic summation is then given by
 \cite{daalhuis1998hyperasymptotic}
 \footnote{$f_{\text{Hyp},0}(w_0)$ is not the same
 as the level-0 hyperasymptotic approximation or optimal
 truncation, since the number of terms at which the series
 is truncated must be increased as more non-perturbative sectors
 are included in the calculation.  
 }
 \begin{equation}
f_\text{Hyp}(w_0) = f_{\text{Hyp},0}(w_0) + \sigma\,f_{\text{Hyp},1}(w_0),
\end{equation}
where the
hyperasymptotic summations for the
perturbative sector and the first non-perturbative sector
are given by 
\begin{align}
	\label{eq:level_one_hyper_f}
	\begin{split}
	f_{\text{Hyp},0}(w_0) =&
	\sum_{m=0}^{N_\text{Hyp}(w_0)-1} \coeff^{(0)}_m
    	w_0^{-m} \\
    	&+ w_0^{1-N_\text{Hyp}(w_0)}
    \frac{S_1}{2 \pi \imu} \sum_{m=0}^{N_\text{Hyp}(w_0)/2-1} \coeff_m^{(1)}
    F^{(1)}\left(w_0;{\genfrac{}{}{0pt}{}{N_\text{Hyp}(w_0)+\beta-m}{-A}}\right); \\
	f_{\text{Hyp},1}(w_0)=&
	\eul^{-A w_0}  w_0^\beta 
    \sum_{m=0}^{N_\text{Hyp}(w_0)/2-1} \coeff^{(1)}_m w_0^{-m};
    	\end{split}
\end{align}
the function $F^{(1)}$ in 
\eqref{eq:level_one_hyper_f} 
is called hyperterminant and
defined in terms of incomplete gamma functions via
\cite{daalhuis1998hyperterminants}
\begin{equation}
	\label{eq:hyperterminant_f1}
F^{(1)}\left(w;{\genfrac{}{}{0pt}{}{M}{a}}\right)  =  \eul^{a w+i \pi  M} w^{M-1}\, \Gamma (M) \Gamma(1-M,a w).
\end{equation}
The quantity $S_1$ in \eqref{eq:level_one_hyper_f} 
is called Stokes constant, which 
may be defined as the change in
the transseries parameter $\sigma$ from
(\ref{eq:transseries_f}) upon crossing the
Stokes line, which in our case is the positive
real axis. The constant $S_1$ has been calculated in previous work \cite{bacsar2015hydrodynamics,aniceto_resurgence_2016} and
 is given by 
\begin{equation}
	\label{eq:stokes_constant}
S_1 \approx 5.4703\times 10^{-3}\,\imu.
\end{equation}
\noindent
This Stokes constant can also be determined using hyperasymptotics, 
see Appendix \ref{ap:stokes_constant}, where we give many more digits.
Note that contributions of order $\bigo{\sigma^2}$ and above
in the transseries \eqref{eq:transseries_f}
are not included in level-one hyperasymptotics.
The error in \eqref{eq:level_one_hyper_f} is therefore
of order $\eul^{-2 |A w_0|}$ \cite{olde2005hyperasymptotics}.

In order to match the late time approximation with the early time solution, we need to bring 
our solution at early times (Eqs. \eqref{eq:f_plus_minus_leading_order} and \eqref{eq:f_c_leading_order})
to the finite value $w_0$. 
This is done by analytical continuation
with the numerical Taylor series method
(see Appendix \ref{ap:taylor}). 
Let us denote the numerical approximation we obtain for $f(w_0)$ as
\footnote{Note that
$f_\text{ac}(w_0)$ depends on which solution we pick around the origin
from the set $\{ f_+, \, f_-, f_C|C\in \mathbb{C}\}$.}
\begin{equation}
    \label{eq:notation_f_an}
    f_\text{ac}(w_0) \quad := \quad 
    \text{numerical analytic continuation of
    $f(w)$ from the origin, evaluated at the time $w_0$} 
\end{equation}

\noindent
By requiring
$f_\text{ac}(w_0) = f_\text{Hyp}(w_0)$, we obtain the following approximation
for $\sigma$:
\begin{equation}
 	\label{eq:sigma_formula_hyper}
 	\sigma \approx \frac{f_\text{ac}(w_0)-f_{\text{Hyp},0}(w_0)}{f_{\text{Hyp},1}(w_0)};
\end{equation}
By decreasing the step size and increasing the order of the Taylor expansions
in the calculation of $f(w_0)$, we can achieve arbitrary accuracy, such that
the error in the approximation \eqref{eq:sigma_formula_hyper}
is determined by limitations of the hyperasymptotic 
approximation. Hence the parameter $\sigma$ in \eqref{eq:sigma_formula_hyper}
is accurate up to an error of order $\eul^{-|A w_0|}$.

\begin{figure}[ht!]
\centering
\includegraphics[width=0.4\linewidth]{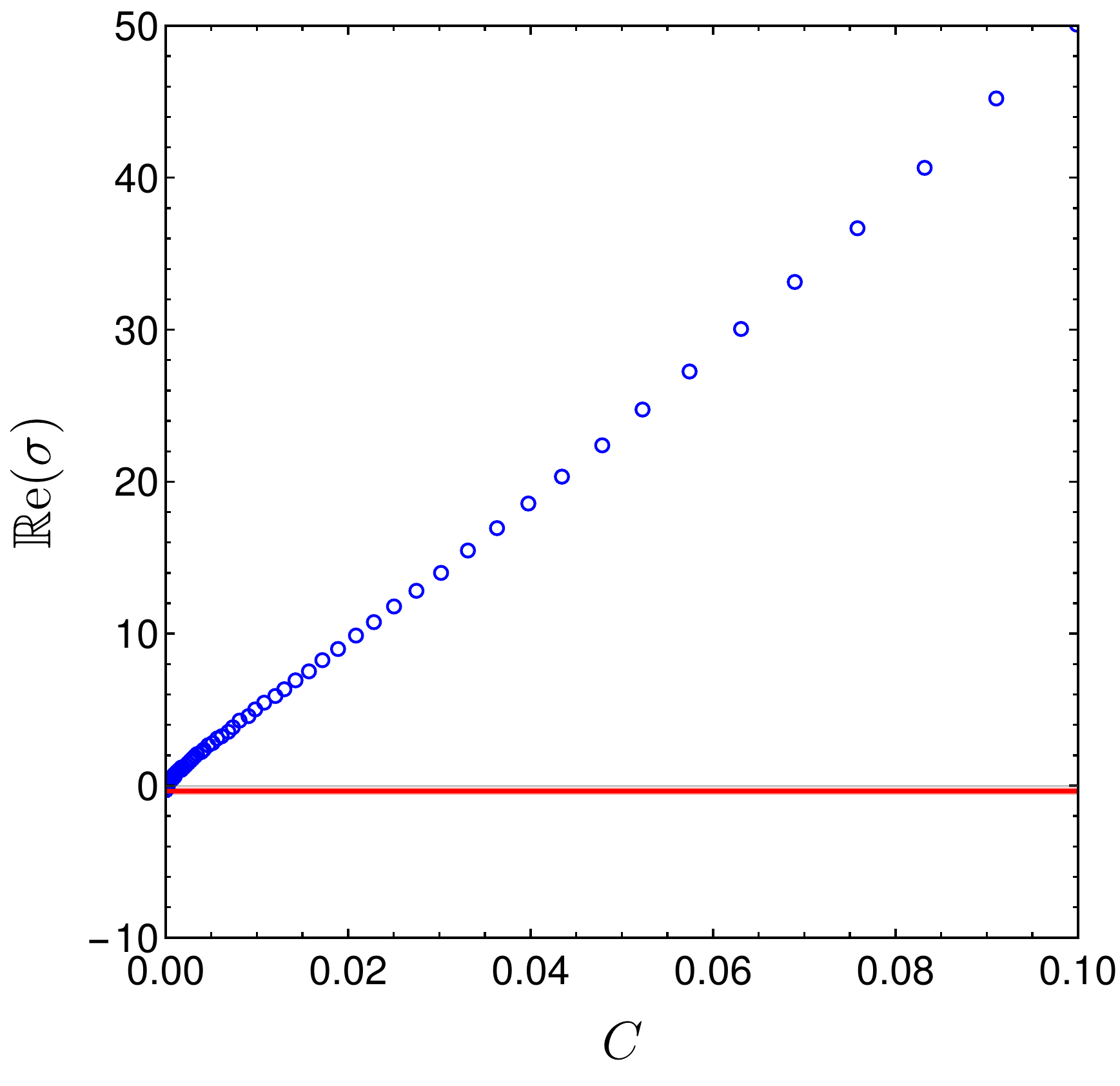}
\qquad\includegraphics[width=0.4\linewidth]{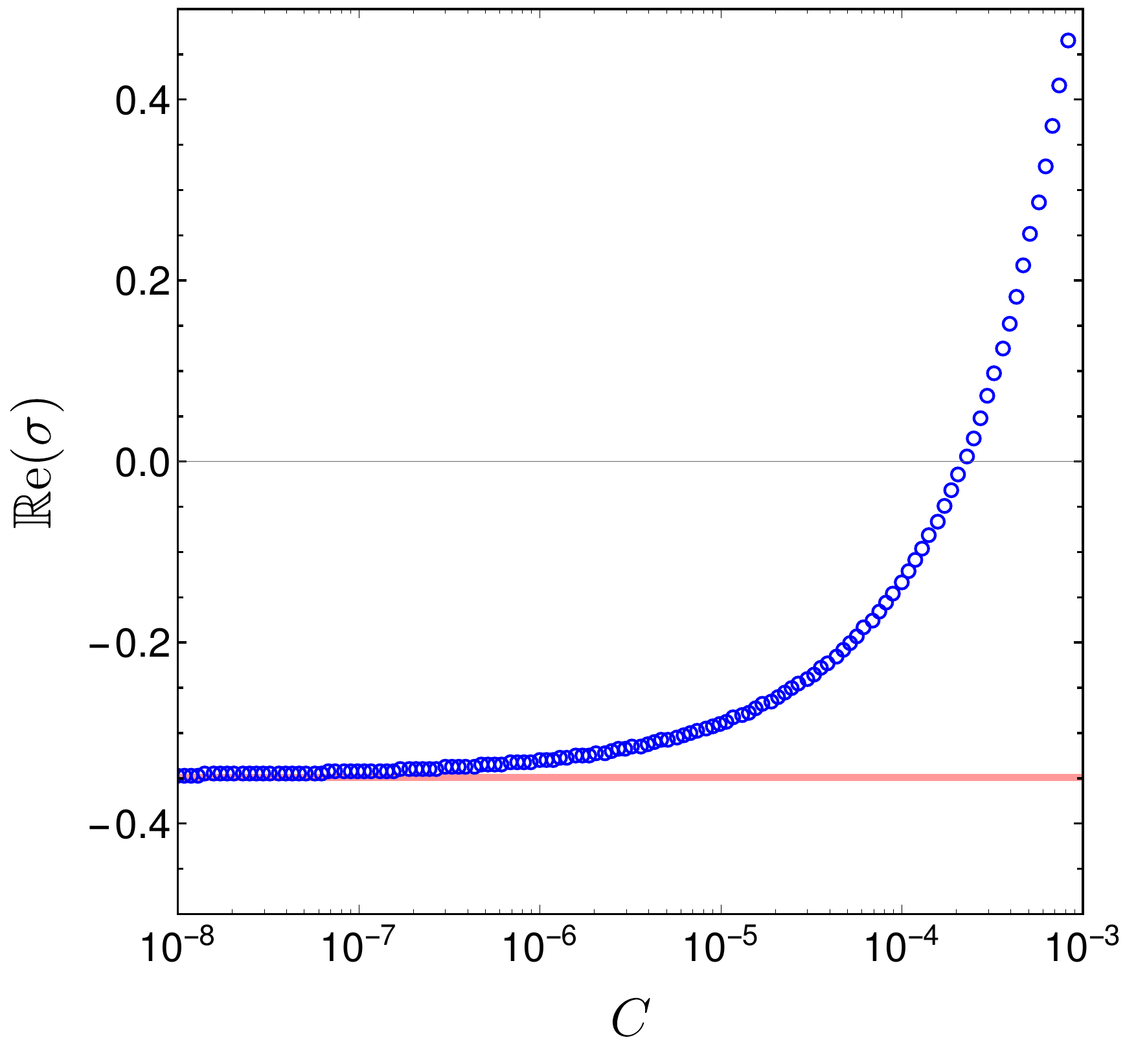}
\caption{The real part of the matched 
late-time transseries parameter $\sigma$
from \eqref{eq:transseries_f} as a
function of the early-time solution $f(w\sim 0)$
displayed 
in a linear plot (left) and a log-linear plot 
(right); Note that the range of values on the
horizontal axis is different in each plot.
Blue dots: matched with solution $f_C(w)$ from
\eqref{eq:f_c_leading_order} for $C>0$;
Red line: matched with solution $f_+(w)$ from 
\eqref{eq:f_plus_minus_leading_order}; For the imaginary
part of $\sigma$ we always have
$\mathbb{I}\text{m}(\sigma) = \mathbb{I}\text{m}(\frac{S_1}{2})$.
The convergence $\sigma(C)\to \sigma_+$ as $C\to0$ shows the
pointwise convergence $f_C(w) \to f_+(w)$.
}
\label{fig:sigma_plot}
\end{figure}

Do note that the approximation for $f(w_0)$ from the late-time transseries solution
can easily be extended to higher orders
in the transseries parameter $\sigma$,
by computing more non-hydrodynamic sectors
$\Phi^{(n)}(w)$ from \eqref{eq:transseries_f}.
In Fig. \ref{fig:sigma_plot} the results of the
early-to-late-time matching $C \leftrightarrow \sigma$
are plotted for $C>0$. Our results are consistent
with the observation in Section
\ref{sec:introduction} that as $(C,\,\sigma) \to (0^+,\sigma_+=-0.349261+0.00273515 \imu )$, the
solutions $f_C(w)$ converge pointwise to the solution
$f_+(w)$, which is finite at the origin.
The function
$\sigma(C)$ in Fig. \ref{fig:sigma_plot} is roughly linear (left plot)
except for a tiny region around
the origin $C=0$, where the converge toward $\sigma_+$ is very slow and
is best visualised on a log-linear plot.

%%%%%%%%%%%%%%%%%%%%%
\subsection{The Borel resummation}
%%%%%%%%%%%%%%%%%%%%%

Another way of approximating $f(w_0)$ is through
Borel resummation (see \textit{e.g.}
\cite{Aniceto:2018bis} for
a review). For a series
$\Phi(w) =\sum_{j\ge0}\,\coeff_j\, w^{-j}$, the
Borel transform of $\Phi(w)$ is given by\footnote{As usual with Borel transforms, any finite number of powers $w^j,\,j\ge 0 $ need to be addressed separately, see \textit{e.g.} \cite{Aniceto:2018bis}.}
\begin{equation}
	\label{eq:borel_transform}
	\borel{\Phi}(\xi)= \coeff_0\,\delta(\xi)+\sum_{j=0}^{+\infty} \frac{\coeff_{j+1}}{j!} \xi^j.
\end{equation}
We truncate the series in
(\ref{eq:borel_transform}) after $N_0$ terms\footnote{
We used $N_0 = 100$, which allows us to perform
the Borel resummation with optimal accuracy up
to $w =7$.} terms and calculate
its Pad\'e approximant $\bpade{\Phi}{N_0}$, \textit{i.e.}
we
approximate the resulting truncated sum by a rational function
$\bpade{\Phi}{N_0}$  with a
numerator/denominator of order $\myfloor{N_0/2}$.

The Borel-Pad\'e resummation method then
consists of taking the inverse Borel transform of
$\bpade{\Phi}{N_0}$, which is given by the
Laplace transform
\begin{equation}
	\label{eq:inverse_borel_pade}
	\mathcal{S}_{N_0,\theta} \Phi(w) = \coeff_0 + \int_0^{\eul^{\imu \theta}\infty} \diffd \xi \,
	\eul^{-w \xi}\bpade{\Phi}{N_0}(\xi).
\end{equation}
The resurgence properties of the transseries \eqref{eq:transseries_f} directly
translates to the existence of singularities of the integrand 
$\bpade{\Phi}{N_0}(\xi)$
in \abbeq \eqref{eq:inverse_borel_pade} 
along the positive real axis -- the Stokes line -- and the singularities reflect the branch cuts of the Borel transform \eqref{eq:borel_transform}, starting at all $\xi=nA,\,\,n\in\mathbb{N}$, one for each exponential in our transseries. Thus
the value of the resummation $\mathcal{S}_{N_0,\theta} \Phi(w)$
depends on the choice of the angle $\theta$ from the positive
real axis. Although this ambiguity in the choice of integration contour gives rise to an imaginary contribution for each summed sector $\mathcal{S}_{N_0,\theta}\Phi^{(n)}(w)$, there is a natural way of summing the resurgent transseries \eqref{eq:transseries_f} such that the final result is unambiguous and real for real positive values of $w$: the so-called \textit{median summation} \cite{Aniceto:2013fka}. To do so we pick a small negative angle $\theta =-\eps <0$
for the integration \eqref{eq:inverse_borel_pade}, and require the imaginary value
of $\sigma$ in the following way (see Appendix \ref{ap:stokes_constant} for some more details):
\begin{equation}
    \label{eq:im_part_sigma_positive}
    \imu \, \mathbb{I}\text{m}(\sigma) =  \frac{S_1}{2}.
\end{equation}

We can now give an 
approximation 
for $f(w_0)$ to first order in the
transseries parameter $\sigma$
\footnote{We are using the transseries 
\eqref{eq:transseries_f} and throwing away
all the terms of order $\bigo{\sigma^2}$ and above.
}:
\begin{equation}
	\label{eq:borel_resummation_f_of_w0}
	f_\text{B}(w_0) \equiv \mathcal{S}_{N_0,-\eps} \Phi^{(0)}(w_0) +\sigma
	w^{\beta}\,\eul^{- A w_0} \mathcal{S}_{N_0,-\eps}  \Phi^{(1)}(w_0);
\end{equation}
In analogy to (\ref{eq:sigma_formula_hyper}), 
we arrive at the following expression for $\sigma$ for the
Borel resummation method:
\begin{equation}
	\label{eq:sigma_formula_borel}
	\sigma \approx \frac{f_\text{ac}(w_0) - \mathcal{S}_{N_0,-\eps} \Phi^{(0)}(w_0)}
	{w^{\beta}\,\eul^{-A w_0}\, \mathcal{S}_{N_0,-\eps} \Phi^{(1)}(w_0)} ;
\end{equation}

Notice that for both \abbeq 
(\ref{eq:sigma_formula_hyper})
and (\ref{eq:sigma_formula_borel}) we only went up to
linear order in $\sigma$ in the approximation of 
$f(w_0)$.
If we wanted to obtain more accurate results, we could 
have
included higher powers of $\sigma$, which amounts to including extra exponential orders\footnote{A similar matching was already done in \cite{heller2015hydrodynamics} for the solution $f_+$ using Borel resummation with two exponentials.}. For the Borel 
summation
method we would only need to numerically compute the 
integrals 
(\ref{eq:inverse_borel_pade}) for the higher-order 
hydrodynamic sectors $\Phi^{(n)}(w)$ in 
\eqref{eq:transseries_f}, while the generalisation of
the hyperasymptotic summation is a bit less straightforward. It 
can nonetheless
be done, and we refer the reader to the literature 
\cite{
olde1995hyperasymptotic,
olde1995hyperasymptotic2,
olde2005hyperasymptotics,
olde2005hyperasymptotics2}.
However, one can obtain the same accuracy if instead of increasing the number of exponentials/powers of sigma, we would just increase the value
of the matching time 
$w_0$.

\begin{figure}[ht]
\centering
\includegraphics[width=0.45\linewidth]{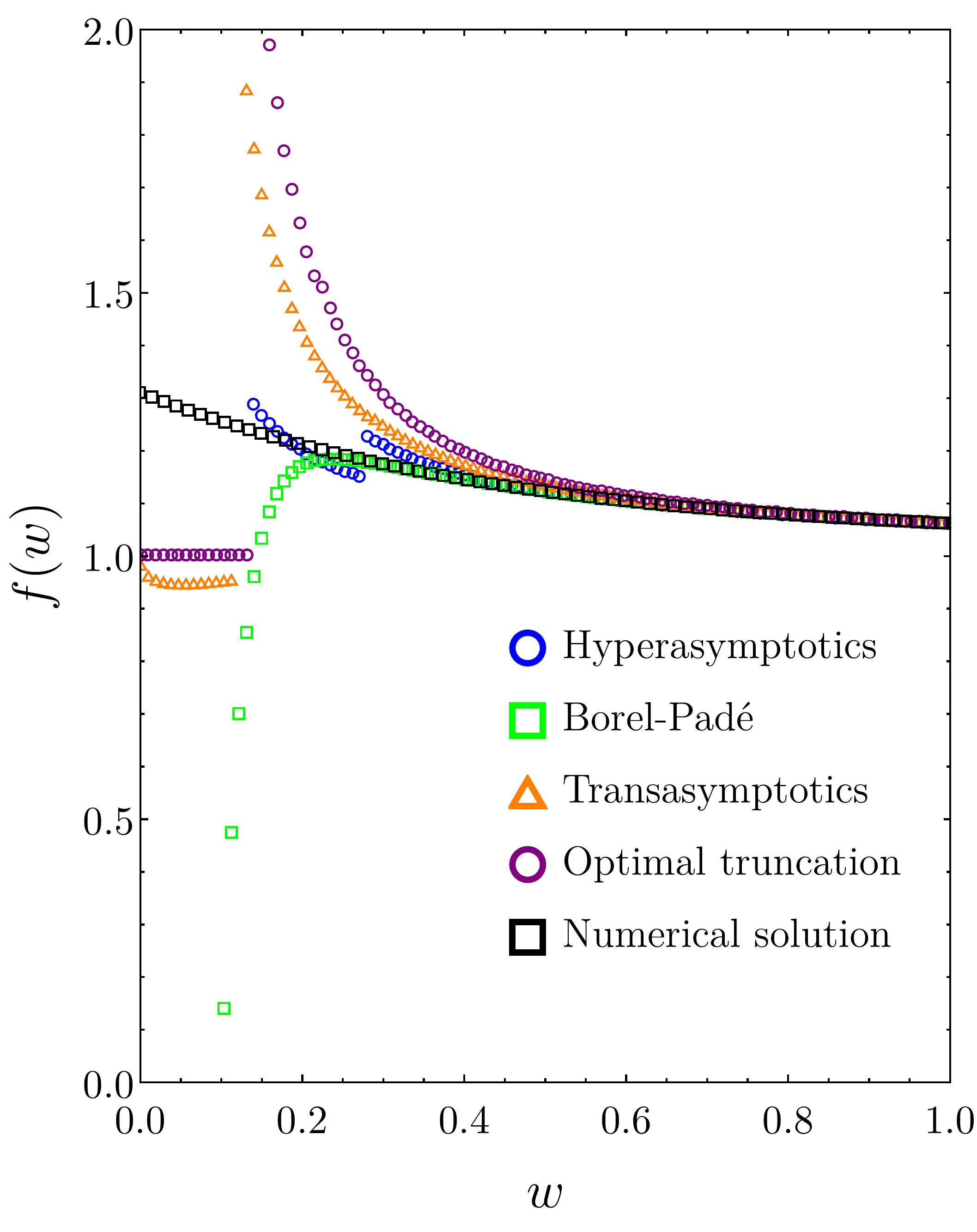}
\includegraphics[width=0.45\linewidth]{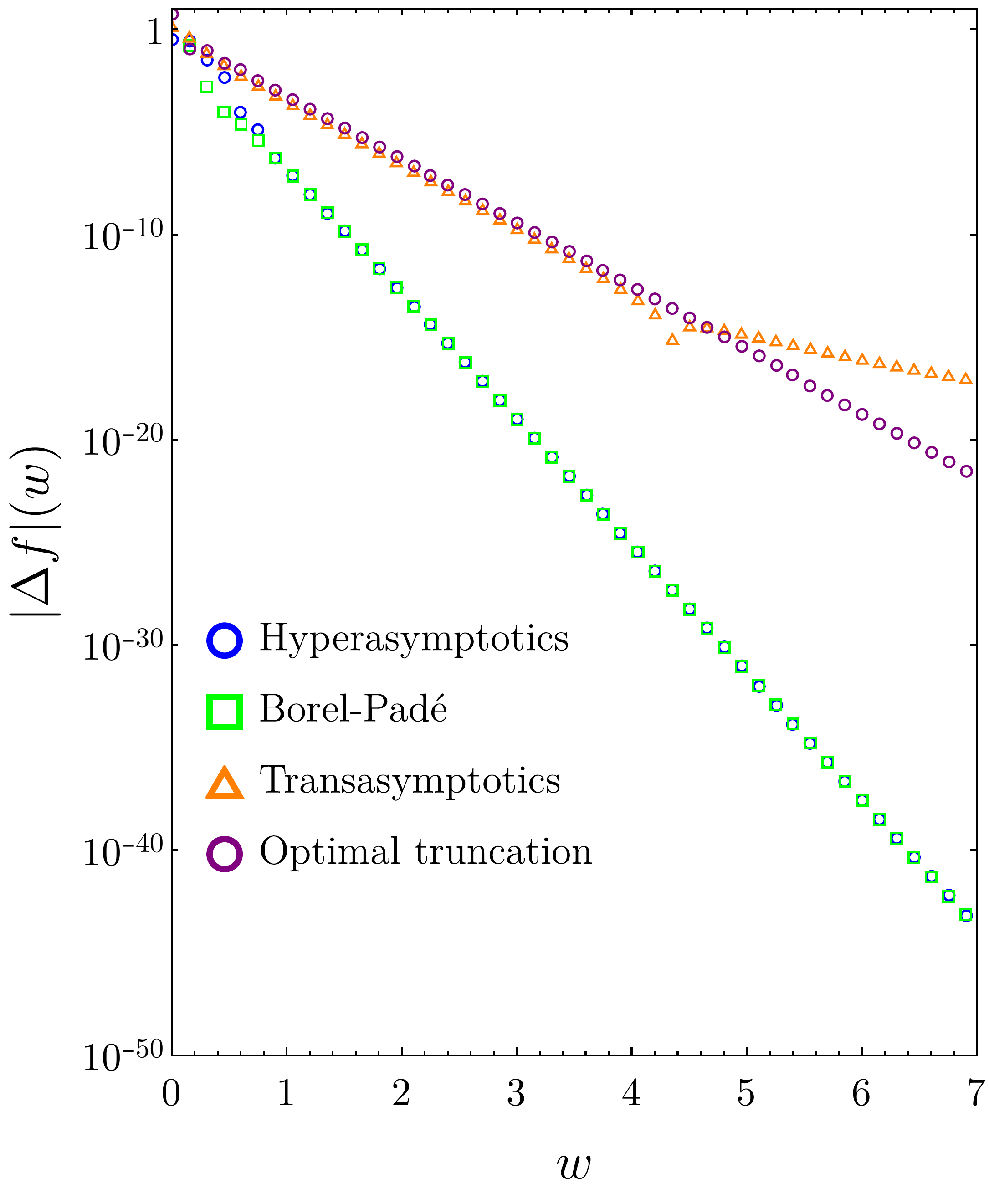}
\caption{Left: Approximations of $f(w)$ using
different resummation methods
 for the transseries parameter $\sigma_+ =-0.3493 + 0.0027 \imu$, 
 corresponding to the function
  $f_+(w)$ (\ref{eq:f_plus_minus_leading_order}). The numerical
  solution is given by the black curve on the left. Right: the absolute value
  of the error of the different methods, which
  has been computed  by comparing the resummations 
  to the numerical solution.}
\label{fig:resummation_plots}
\end{figure}

Once the parameter $\sigma$ has been matched to a given initial
condition,\footnote{Value of the function  at $w=0$ for the solution $f_+(0)$, or for the solutions which
diverge at the origin 
$f_C(w)\sim C w^{-4}$, the value of $C$ is used as an
initial value.} the transseries (\ref{eq:transseries_f}) can be
used to find an approximation of $f(w)$ everywhere, 
via some summation technique such as hyerasymptotics and Borel summation described above.
The hyperasymptotic method does not require
computing numerical
integrals, but has the disadvantage of yielding
discontinuous approximations to the summed transseries: it provides a piecewise analytic approximation (which is clear from the left plot of Fig. \ref{fig:resummation_plots}).
On the other hand, 
the Borel summation integrals 
(\ref{eq:inverse_borel_pade}) must be computed as numerical
approximations at each evaluation point, but the method has the advantage of
giving a continuous function of $w_0$. 

In Fig.
\ref{fig:resummation_plots}, we can see how 
different resummation methods compare with 
each other: in terms
of accuracy the hyperasymptotic summation and
the Borel resummation method are equivalent
outside of a very small region near the origin,
both giving an exponentially small error of
approximately $\sim \eul^{-2|A w|}$ (the order of the first 
exponential we have neglected). We can also clearly see that the approximations given by each summation method are quite accurate at very early times even though we have only included a single exponential mode -- to obtain accurate results at earlier times one would need to include further exponentials and their respective asymptotic expansions from \eqref{eq:transseries_f}.

Also in Fig. \ref{fig:resummation_plots} one can find results corresponding to a transasymptotic 
resummation, which will
be discussed in the next Section \ref{sec:transasymptotics}. Let us also briefly
mention the optimal truncation method, which consists
of truncating the power series of the perturbative sector before the least term \footnote{The formula 
for $N_\text{opt}$ in
\eqref{eq:pert_sector_power_series}
is a good approximation
for the least term.
}
\begin{equation}
    \label{eq:pert_sector_power_series}
    f_\text{opt}(w) = \sum_{n=0}^{N_\text{opt}(w)-1}\,\coeff^{(0)}_n
    w^{-n},\quad \text{where}\qquad  N_\text{opt}(w) = \myfloor{| A w |}.
\end{equation}
The accuracy of the optimal truncation method
is approximately $\sim \eul^{-|A w|}$, which 
agrees with our plots in Fig.\ref{fig:resummation_plots}.
Now that we have discussed how to perform
the interpolation between late and early times
using
Borel resummation and hyperasymptotics, 
the next 
section will be devoted to
the transasymptotic summation method.

%%%%%%%%%%%%%%%%%%%%%%%%%%%%%%%%%%
\section{Transasymptotic summation}
	\label{sec:transasymptotics}
%%%%%%%%%%%%%%%%%%%%%%%%%%%%%%%%%%

We have seen in Section \ref{sec:late_and_early_times}
that approximating the transseries
\eqref{eq:transseries_f} by keeping only
the perturbative and the first non-perturbative
sector gives excellent approximations
of exponential accuracy for the function $f(w)$
outside a small region near the origin. However,
truncating the transseries in this way only
works if the exponentials are small. Along the
negative axis,
the exponential monomial
$\tau \sim \eul^{-A w}$ 
defined in \eqref{eq:def_tau} grows
arbitrarily large, and it is clear that
truncating the transseries 
\eqref{eq:transseries_f} can no longer work
since all orders of $\tau$ contribute significantly
towards the sum in that regime.
This raises the question whether the transseries
is of any use at all in regions where the
exponential monomial is large enough. 
The answer is yes: we can 
exploit the fact that the divergent behaviour in
the transseries comes only from large orders of
the perturbative variable $w^{-1}$, whereas
the large order behaviour of 
the exponential variable
$\tau$ is convergent. All we need to do
is change the order of summation in 
\eqref{eq:transseries_f}:
\begin{equation}
     \label{eq:transas_summation}
    \mathcal{F}(\tau,w)=\sum_{n \geq 0} \sum_{r \geq 0} \tau^n \coeff^{(n)}_r w^{-r} 
    = \sum_{r\geq 0} \left(\sum_{n\geq 0} \coeff^{(n)}_r
    \tau^n \right) w^{-r} 
    \equiv  \sum_{r\geq 0} F_r(\tau)  w^{-r};
\end{equation}
The coefficient functions $F_r(\tau)$ are 
analytic at $\tau =0$ , and we will
see that it is possible to systematically
calculate them in closed form. This approach
is called the transasymptotic summation \cite{costin1999correlation,costin2001formation}, and
has been shown to be a powerful tool in the
study of non-linear problems 
\cite{costin2015tronquee,aniceto2021capturing,behtash2021transasymptotics}.

The special form of (\ref{eq:transas_summation}) allows us
to compute the functions $F_r(\tau)$ by treating $\tau$ and $w$ as 
independent variables. Let us start with the lowest order
approximation
\begin{equation}
	\label{eq:transasymptotic_f_lowest_order}
	\mathcal{F}(\tau,w) = F_0(\tau) + \bigo{w^{-1}}, \quad w \to \infty
\end{equation}
Then $F_0$ obeys the ODE
\begin{equation}
	\label{eq:ode_f0}
	-1 + F_0(\tau) \left( 1- \tau F_0'(\tau)\right)= 0,
\end{equation}
which is solved by $F_0(\tau)=1+W(\frac32\tau)$
\footnote{The general solution to \eqref{eq:ode_f0} is
$F_0(\tau) = 1+W(c \tau)$. The integration constant $c$
is found by matching the transasymptotic expansion
to the transseries \eqref{eq:transseries_f}, and depends
on the choice for $\coeff^{(1)}_0$. 
Our choice is $\coeff^{(1)}_0=3/2$.}, where $W$ stands
for the branch $W_0$ of the Lambert-W function (see Appendix \ref{ap:lambert_w}). 
We can go further and
calculate all $F_r(\tau)$. For $r \geq1$ we find the following
differential equations for $F_r$:
\begin{align}
	\label{eq:recursion_f_r}
	\begin{split}
	A\left(\tau F_0(\tau)F'_r(\tau)+\left(\tau F'_0(\tau)-1\right)F_r(\tau)\right)
	=&(4-\beta) \delta_{r,1} - 8 F_{r-1}(\tau)
	+ \frac{9-r}2 \sum_{k=0}^{r-1} F_k(\tau)F_{r-1-k}(\tau)\\
	&+\beta\tau \sum_{k=0}^{r-1} F_k(\tau)F'_{r-1-k}(\tau)
	-A\tau \sum_{k=1}^{r-1} F_k(\tau)F'_{r-k}(\tau);
	 \end{split}
\end{align}
Note that in (\ref{eq:recursion_f_r}) 
all the derivative terms come multiplied by the variable $\tau$,
and that the variable $\tau$ does not appear other than
as a multiplier of the derivatives. This motivates the convenient
variable transformation $\tau \to W = W(\frac32\tau)$. The derivatives
transform as 
\begin{equation}
	\label{eq:derivative_w_identity}
	\tau \diff{}{\tau} = \frac{W}{1+W} \diff{}{W} .%\equiv \Theta(W) \diff{}{W},
\end{equation}

With the transformation (\ref{eq:derivative_w_identity}) it is
possible to rewrite the original recursive set of ODEs (\ref{eq:recursion_f_r})
and integrate them exactly. The details of this calculation as well as
the method of fixing the integration constants are given in
Appendix (\ref{ap:calculation_f_r}). It turns out that all the $F_r$ are
rational functions in $W$ and can be computed exactly \cite{StringMathtalk,Borinsky:2020vae}.
Let us now see how the functions $F_r(\tau)$ can
be used to solve the interpolation problem
between early and late times.

%%%%%%%%%%%%%%%%%%%%%%%%%%%%%%%%%%%%%%%%
\subsection{Interpolation with transasymptotics}
%%%%%%%%%%%%%%%%%%%%%%%%%%%%%%%%%%%%%%%%%

We want to find an approximation for
the transseries parameter $\sigma$ corresponding to a given
solution around the origin (\eqref{eq:f_c_leading_order}
or \eqref{eq:f_plus_minus_leading_order}) using 
the transasymptotic summation 
\eqref{eq:transas_summation}. The first step of
our approach is the same as in 
Section \ref{sec:late_and_early_times}: we use
numerical analytical continuation
from the origin to the
matching point $w=w_0$, obtaining the numerical
approximation $f_\text{ac}(w_0)$ 
(see \eqref{eq:notation_f_an}). In a second
step,
we compute an approximation for
$\tau(w_0)$, from which
 the transseries parameter
$\sigma$ can directly be 
calculated
using our definition of $\tau(w_0)$, \abbeq \eqref{eq:def_tau}. The idea
is the following: we want to solve for
the function $\gamma(w)$ 
obeying
\begin{equation}
	\label{eq:transasymptotics_is_constant} 
 \mathcal{F}(\gamma(w), w) =
  \sum_{n \geq 0}\, F_r\left(\gamma(w)\right) w^{-r} =f_\text{ac}(w_0)= \text{constant},
  \text{      for all } w,
\end{equation}
which will be equal to $\tau(w_0)$ when
evaluated at the point $w_0$, \textit{i.e.} 
$\gamma(w_0)= \tau(w_0)$.
The function $\gamma(w)$ satisfying
\eqref{eq:transasymptotics_is_constant}
admits a perturbative, divergent 
asymptotic expansion in $w^{-1}$:
\begin{equation}
\label{eq:gamma_zeroth_ord}
\gamma(w) = \sum_{k=0}^{+\infty}\, \gamma_k\,w^{-k}\,,%\gamma_0 + \bigo{w^{-1}}, \text{as } w \to \infty.
\end{equation}

\noindent
and determining $\gamma(w)$ will correspond to finding the coefficients $\gamma_k$. Truncating the above expansion at its first term $\gamma(w)=\gamma_0+\bigo{w^{-1}}$, we find from \eqref{eq:transasymptotics_is_constant} that up to leading order
\begin{equation}
    1+W\left(\frac{3}{2}\gamma_0\right)=f_{\text{an}}(w_0).
\end{equation}
 Then also up to leading order in $w_0^{-1}$, we have $\gamma_0=\tau(w_0)$, which together with the definition of $\tau(w)$ \eqref{eq:def_tau} returns:
\begin{equation}
 	\label{eq:gamma_of_sigma_lowest_order}
 	 \sigma(w_0) = \big(f_\text{ac}(w_0)-1\big) \,w_0^{-\beta}\,
 	 \mathrm{e}^{ f_\text{ac}(w_0)-1 +A w_0} \bigg(1+ \bigo{w_0^{-1}} \bigg).
 \end{equation}
\abbeq \eqref{eq:gamma_of_sigma_lowest_order}
can be easily extended to higher 
orders in $w_0^{-1}$ by including higher
orders in the ansatz  \eqref{eq:gamma_zeroth_ord}
and matching powers of $w^{-1}$ in
\eqref{eq:transasymptotics_is_constant}.
The first four coefficients of the perturbative expansion
of $\gamma(w)$ are given in Appendix \ref{ap:gamma_perturbative}.
The transasymptotic summation \eqref{eq:transas_summation}
can also be used to re-sum the transseries by truncating the
series at the term of least magnitude. The difference with respect
to the classical optimal truncation is that coefficients $F_r(\tau(w))$
vary with $w$. The result is displayed in Fig. \ref{fig:resummation_plots}.
We can see that this approach outperforms optimal truncation. Note that we only calculated the coefficient functions $F_r(w)$ up to $r=15$,
and so the calculation is no longer optimal after the kink
in the logarithmic error plot of Fig. \ref{fig:resummation_plots}. Furthermore, the
kink happens at a higher value of $w$ than
we would expect from the resummation
point $w_0$ corresponding
to $15$ terms with classical optimal truncation
given by \eqref{eq:pert_sector_power_series}.

%%%%%%%%%%%%%%%%%%%%%%%%%%%%%%%%%%%%%%%%%%%%%
\subsection{Analytic results: branch points and global behaviour}
%%%%%%%%%%%%%%%%%%%%%%%%%%%%%%%%%%%%%%%%%%%%%

Transasymptotics can be used to describe \textit{global properties}
of the function $f(w)$ 
from \eqref{eq:ode_mis_model_f_of_w}, 
such as zeros, poles, branch points or to link distinct
expansions in different asymptotic regimes. This 
is quite remarkable given that the transasymptotic
summation was derived as a \textit{local expansion}
around the point $w=+\infty$. 
Let us start by
sketching out how the locations of the branch points
may be obtained.
Notice that in the solutions 
plot Fig. \ref{fig:regularzero} the
locations $w_\text{s}$ of the square root 
branch points depend on 
the initial value problem
that $f(w)$ solves. 
From the perspective of late-time asymptotics,
this means
that the locations $w_\text{s}$ 
are a function of the transseries
parameter $\sigma$. 
As already mentioned, 
all the coefficient functions $F_r(\tau)$ in
the transasymptotic summation (\ref{eq:transas_summation})
can be expressed as rational 
functions of the Lambert-W function $W(\frac{3}{2}\tau)$,
which has a square-root branch point 
at $\tau=-\frac{2}{3}\eul^{-1}$. This branch point
in the $\tau$-plane translates to an infinite
number of branch points in the $w$-plane
if we substitute $\tau=\tau(w)$ from \eqref{eq:def_tau}.
Since the Lambert-W function
appears in all the coefficient functions
in the transasymptotic summation
\eqref{eq:transas_summation}, 
we expect the function $f(w)$
to have an infinite number of square root branch points
as well. The analytic information about
the non-perturbative exponentials encoded in the coefficient functions
$F_r(\tau)$ can be used 
to provide an approximation for the locations $w_\text{s}$.

Do note that all zeros of $f(w)$
 are square root branch point singularities 
 (see the expansions in  \abbeq \eqref{eq:f_of_w_branch_point_sols})
with the exception of a potential regular zero at
 $w= (4-\beta)/A \approx 0.502$. 
 Hence we can
solve for the branch points $w_\text{s}$ by solving the equation
\begin{equation}
\label{eq:condition_transas_zero}
    \mathcal{F}(w=w_\text{s},\sigma) = 0
\end{equation}
approximately 
for $w_\text{s}(\sigma)$, where $\mathcal{F}$ is the
transasymptotic summation from \abbeq (\ref{eq:transas_summation}). 
We find
\begin{equation}
	\label{eq:movable_singularities_equation}
  w_\text{s}(t) \simeq w^\text{(approx)}_\text{s}(t)= \frac{t}{A}+\frac{\beta}{A}\, \log t+\frac{1}{A\, t}\left(\beta^2\,\log t+\beta^2-5\beta-\frac{3}{A}\right)\, , \quad \text{   as } t \to \infty,
\end{equation}
where we have introduced the following variable:
\begin{equation}
	\label{eq:t_variable_expression}
	t\equiv t(n,\sigma) = \log\left(\frac{3\sigma \eul}{2A^\beta}\right) + \pi \imu (1+2n ), \quad \quad n \in \mathbb{Z};
\end{equation}
The integer $n$ in (\ref{eq:t_variable_expression}) parameterises
the sequence of branch points. 
Note that (\ref{eq:movable_singularities_equation}) is the
partial sum of a
divergent 
asymptotic expansion in $t$, and thus \abbeq \eqref{eq:movable_singularities_equation}
is only a good approximation for the branch points/zeros of $f(w)$ when
$|t|$ is large. In particular, $w^\text{(approx)}(t)$ becomes more accurate
for large values of the discrete parameter $n$ from
 \eqref{eq:t_variable_expression},
 since the
 auxiliary variable $t$ grows as an affine function with $n$.  Since the leading
 order approximation $w^\text{(approx)}(t)$ from \abbeq 
 \eqref{eq:movable_singularities_equation} grows linearly in $t$, 
the branch points which lie far from the origin are
 are best approximated by $w^\text{(approx)}(t)$.

  \begin{figure}[ht]
\centering
\includegraphics[width=0.6\linewidth]{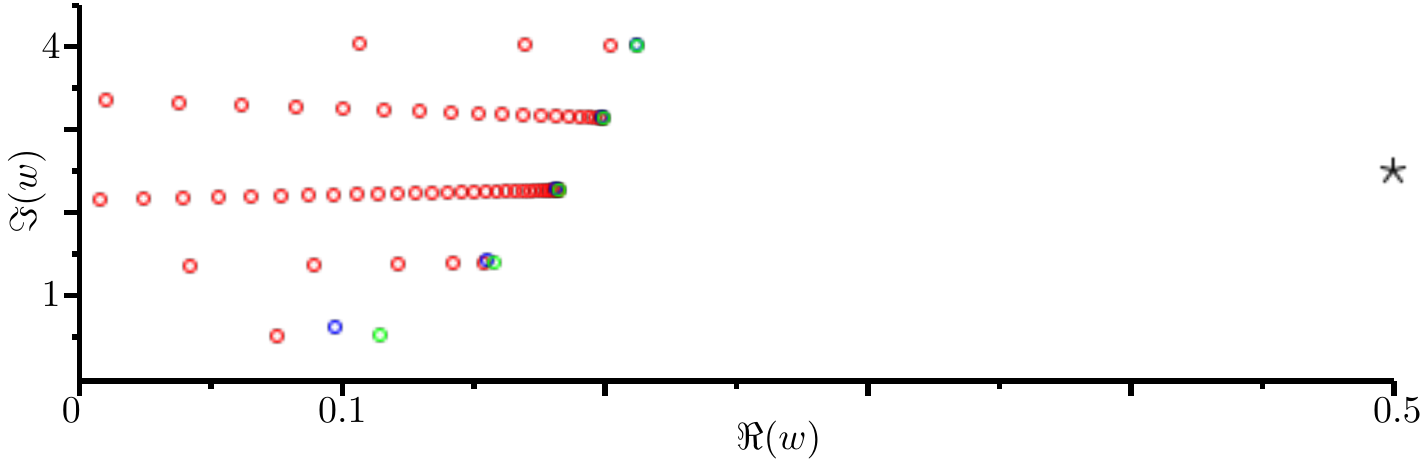}
\caption{Branch cuts of the $f(w)$ for the case $\sigma=\frac23$. Green dots: numerically computed branch-points; blue dots:
 approximations  
 \eqref{eq:movable_singularities_equation}
 to the locations of the branch points
 obtained from the transasymptotic summation of the late time solution $\mathcal{F}(w)$ \eqref{eq:transseries_f} (compare Table \ref{tab:branch_points}).
Red dots: poles for the Pad\'e approximant (about $w=\frac12+\frac52\imu$, shown as $\star$) representing the branch cuts of the solution.}
\label{fig:branchpoints}
\end{figure}

Numerically, we can compute the zeros  of  
$f(z)$ by initially guessing
the position of the branch point using \eqref{eq:movable_singularities_equation}
\footnote{Note that we could also 
have used Pad\'{e} approximants for the initial guess.}
and then using a contour integral to find a good approximation for the
exact location.
We start by choosing a value for the transseries
parameter $\sigma$ and use
 the hyperasymptotic approximation  \abbeq \eqref{eq:level_one_hyper_f}
  to find $f(w_0)$ (\textit{e.g.} $w_0=10$). We then analytically
continue $f(w)$ from $w_0$ to a point in the vicinity of
our prediction  \eqref{eq:movable_singularities_equation} 
using the Taylor series method, $w_1 = w^\text{(approx)}_\text{s}(t) + \varepsilon$ 
(\textit{e.g.} $\varepsilon = 0.3$). Next we analytically continue again
to compute the data on the circle 
$|w- w^\text{(approx)}_\text{s}(t)| =\varepsilon$. Using the trapezoidal rule
 \cite{TW14} we evaluate the contour integral of $\frac{w f^\prime(w)}{f(w)}$
 to obtain the zeros of $f(w)$
\footnote{Due to the square root singularity
 (see \eqref{eq:movable_singularities_equation}),
the branch point must be encircled twice.}. The approximate locations
obtained with \abbeq \eqref{eq:movable_singularities_equation}
as well as the numerical results 
are listed in Table \ref{tab:branch_points},
and plotted in
 Fig. \ref{fig:branchpoints}.

\begin{table}[ht]
\begin{center}
\begin{tabular}{|c|c|c|}
\hline
 & approx.~\eqref{eq:movable_singularities_equation} 
 & numerical \\
\hline
$n=0$
& $0.0975 + 0.6040\imu$ 
& $0.1147 + 0.5076\imu$\\

$n=1$
& $0.1555 + 1.416\imu$
& $0.1580 + 1.384\imu$\\

$n=2$
& $0.1817 + 2.276\imu$
& $0.1827 + 2.257\imu$\\

$n=3$
& $0.1991 + 3.143\imu$
& $0.1997 + 3.129\imu$\\

$n=4$
& $0.2122 + 4.012\imu$ 
& $0.2125 + 4.001\imu$ \\

\hline
\end{tabular}
\caption{Approximations for the
locations of the square-root
branch points of
\eqref{eq:movable_singularities_equation}
versus their numerically computed values for
$\sigma=\frac23$.}
\label{tab:branch_points}
\end{center}
\end{table}%

Let us now turn to another
powerful application of transasymptotics:
it can be 
used to correctly predict the different asymptotic 
behaviour of our solutions in separate
regions. Consider the attractor 
$f_+$ in the 
solutions plot Fig. \ref{fig:regularzero} 
(the black curve).  
At large, positive $w$,
$f_+(w)$ converges to a finite value, 
while at large, negative $w$
the same solution grows linearly with $w$. 
Therefore we
have have two different asymptotic expansions,
 the transseries (\ref{eq:transseries_f}) at
 large positive $w$ and
the linearly growing expansion (\ref{eq:p4_family_solutions}) 
at large negative $w$
(which is also a transseries, 
but with $\log w$-monomials
instead of exponentials $\eul^{-A w}$, 
see \cite{edgar2010transseries}).
This is not surprising given the presence of square root branch points
in the domain of our solutions. But it also raises an interesting
question:
can we somehow relate the two expansions to one another?
The answer is yes: the great power of the transasymptotic
approach lies in the possibility of analytically accessing 
regions in which the non-perturbative exponentials are
no longer small. While the large, positive $w$ limit corresponds
to exponentially small values of $\tau\sim \eul^{-A w}$, 
the large, negative $w$ limit is 
associated with exponentially large values of $\tau$. Since
the coefficient functions in the transasymptotic summation 
\eqref{eq:transas_summation} are just rational functions
of $W(\tau)$, and the large $\tau$ expansion of $W(\tau)$
is known 
(see \cite{corless_lambertw_1996} and Appendix \ref{ap:lambert_w}),
we were able to use transasymptotics to correctly derive
the first four terms of the other, linearly growing expansion
\eqref{eq:p4_family_solutions}.
The reason transasymptotics is so powerful in this
case is that in flipping the sign 
 $w \to -w$, the powers
of $w^{-1}$ in the transasymptotic summation do not
change size, while the exponential
variable $\tau\sim \eul^{-A w}$ changes its regime, and 
becomes exponentially large instead of exponentially
small. The details of the
calculations in this section
are beyond the scope of this publication and will be explored in 
an upcoming paper \cite{upcoming_paper}.

%%%%%%%%%%%%%%%%%%%%%%%%%%%%%%%%%%%%%%%%%%%%%%%%%
\section{Summary/Discussion}
\label{sec:summary}
%%%%%%%%%%%%%%%%%%%%%%%%%%%%%%%%%%%%%%%%%%%%%%%%

The main focus of this work was to solve the problem
of late time to early-time matching for arbitrary solutions
of the ODE (\ref{eq:ode_mis_model_f_of_w}). 
We  have a one-parameter family
of solutions in two different regions of our domain: at late times, 
the ODE (\ref{eq:ode_mis_model_f_of_w})
admits formal transseries solutions consisting of the hydrodynamic
perturbative sector as well as non-hydrodynamic sectors incorporating
positive integer powers of the non-perturbative exponentials
 $\eul^{-A w}$ in the variable $w$ representing time.
In the early time regime near $w \sim 0$
there is a one-parameter family of divergent
solutions which behave asymptotically as $\sim w^{-4}$,
as well as two finite solutions which are 
special limits of the one-parameter family (see solution plots
Fig. \ref{fig:regularzero}). 
The different exponentially small contributions 
appearing at late times can be expected to be 
the leading contributions at early times. Beyond the MIS case, one expects to find similar transseries solutions in other hydrodynamic systems which observe a factorially divergent late time behaviour (see \textit{e.g.} \cite{Heller:2021yjh})

The resummation
methods we used are hyperasymptotics,
Borel-summation, and transasymptotics, and
are all well-established.
However, previous work did not exploit their strengths
to do the parameter-matching and relied instead on less accurate 
procedures such as numerical 
least square fits \cite{behtash2021transasymptotics,
Behtash:2019txb}.
We carried out an analysis of said methods, and have shown that they
are very effective tools for the parameter-matching.
In terms of accuracy, the hyperasymptotic approach and the Borel
resummation perform best. Both give an exponentially small error
$\sim \eul^{-2| A w|}$ in the variable $w$. 

The hyperasymptotic approximation has
discontinuities since the number of terms which are included in
the series varies with $w$, but requires no intricate 
numerical computations
other than determining
the series coefficients of the perturbative and
first non-perturbative sectors. 

On the other hand,
the Borel resummation is a continuous function of $w$. Both resummation methods can be extended
to include an arbitrary number of 
exponentials.
Hence the method can be made arbitrarily accurate by increasing the
number of non-perturbative modes 
we include in the approximation.
However,
the calculation of the Laplace transform
(\ref{eq:inverse_borel_pade}) in going from the Borel-plane to
the complex plane of our original variable $w$ requires
the numerical computation of an integral. As a consequence,
Borel resummation is more computationally
expensive than the 
hyperasymptotic summation, especially since
said integral must be computed to exponential accuracy $\eul^{-A w}$
 for the method to perform
as well as the hyperasymptotic summation. 

The transasymptotic summation
has originally been used very effectively
in the analysis of solutions of non-linear problems 
\cite{costin2001formation}.
While the transasymptotic 
summation is less accurate in 
performing the interpolation, giving an error of $\sim \eul^{-|A w|}$
(as opposed to $\eul^{-2 |A w|}$ for the other methods),
it is an extremely
useful tool in the study of the \textit{global} analytic properties of
the solutions. 
The power of the transasymptotic approach
lies in encoding
the behaviour of the non-perturbative exponentials in analytic
closed-form expressions, the transasymptotic coefficient functions. 
We have provided a systematic way of calculating these
functions and used them to derive intricate analytic results such as
analytic approximations to the locations of the square-root
branch points as well as a way of linking distinct asymptotic expansions
in two different regions of the domain to each other. These results
have only been sketched out in this work, and a larger analysis will
be presented in an upcoming paper 
\cite{upcoming_paper}.

The matching procedure we used is quite general and can 
be used beyond relativistic hydrodynamics. In fact, we can apply it to
any interpolation
problem between two different
regions (e.g. late-time to early-time, strong/weak coupling, large charge to small charge),
where the solutions in one region are described by
 resurgent, asymptotic perturbative expansions, and where the behaviour
 in the other
regime is known analytically (\textit{e.g.}
\cite{heller2018does,
behtash2020global,
behtash2021transasymptotics,
aniceto2016resurgencecusp,Dorigoni:2015dha,Romatschke:2017acs,
Du:2022bel}).

%%%%%%%%%%%%%%%%%%%%%%%%%%%%%%%%%%%%%%%%%%%%%%%%%
\section*{Acknowledgements}
%%%%%%%%%%%%%%%%%%%%%%%%%%%%%%%%%%%%%%%%%%%%%%%%

The authors would like to thank the participants of the focus week on Relativistic hydrodynamics during the programme \textit{Applicable Resurgent Asymptotics} at the Isaac Newton Institute for the many relevant discussions that took place, and Ben Withers for his feedback on a draft of this work.  The authors would also like to thank the Isaac Newton Institute for hosting them during the early stages of the work. IA has been supported by the UK EPSRC Early Career Fellowship EP/S004076/1, and the FCT-Portugal grant PTDC/MAT-OUT/28784/2017. DH has been supported by the presidential scholarship of the University of Southampton.

\appendix

%%%%%%%%%%%%%%%%%%%%%%%%%%%%%%%%%%%%%%%
\section{Recurrence relations for \texorpdfstring{$\Phi^{(1)}$}{phi1} and \texorpdfstring{$\Phi^{(2)}$}{phi2}}
\label{ap:recurrence_relations}
%%%%%%%%%%%%%%%%%%%%%%%%%%%%%%%%%%%%%%%

The recurrence relations for the coefficients of the perturbative
and the first non-perturbative sector can be derived by substituting
the expression
\begin{equation}
\label{eq:transseries_linear_order}
f(w) = \Phi^{(0)}(w)  + \sigma w^\beta \eul^{-A w} \Phi^{(1)}(w)
\end{equation}
into the MIS ODE (\ref{eq:ode_mis_model_f_of_w}). At order $\bigo{\sigma^0}$
we obtain the same ODE, but for $\Phi^{(0)}$ instead of $f$. At order
$\bigo{\sigma^1}$ we find the equation
\begin{equation}
\label{eq:order_sigma_equation}
\Phi^{(1)}(w) \left( -8 + A w + \left(8-A w+ \beta\right) \Phi^{(0)}(w) 
+w \partial_w \Phi^{(0)}(w)\right) + w \Phi^{(0)}(w) \partial_w \Phi^{(1)}(w) =0.
\end{equation}
With the series ansatz
\begin{equation}
\Phi^{(n)}(w) = \sum_{j=0}^{\infty} \coeff^{(n)}_j w^{-j},
\end{equation}
we obtain the recurrence relations for $\Phi^{(0)}$ and $\Phi^{(1)}$:
\begin{align}
\label{eq:rec_rel_phi0}
\begin{split}
\coeff^{(0)}_0 =& \,1; \quad \coeff^{(0)}_1 = \frac{\beta}{A}; \\
\coeff^{(0)}_{j} =& \frac{1}{A} \left[ 
8\coeff^{(0)}_{j-1}+\frac{j-9}{2}\sum_{\ell=0}^{j-1} \coeff^{(0)}_\ell \coeff^{(0)}_{j-1-\ell}\right], \quad \text{for } j\geq2; \\
\coeff^{(1)}_0 \equiv& \, \frac32; \\
\coeff^{(1)}_{j} =& \frac{1}{j} \left[ 
(8+\beta-j) \sum_{\ell=0}^{j-1}\coeff^{(1)}_\ell \coeff^{(0)}_{j-\ell} 
- A \sum_{\ell=0}^{j-1} \coeff^{(1)}_\ell \coeff^{(0)}_{j+1-\ell} \right], 
\quad \text{for } j\geq1;
\end{split}
\end{align}
The coefficient $\coeff^{(1)}_0$ is undetermined by (\ref{eq:order_sigma_equation}),
and any redefinition of $\coeff^{(1)}_0$ can be absorbed into the
transseries parameter $\sigma$.

%%%%%%%%%%%%%%%%%%%%%%%%%%%%%%%%%%%%%%%%%%%%%%%%%%
\section{The Stokes constant \texorpdfstring{$S_1$}{S1} and median summation}
\label{ap:stokes_constant}
%%%%%%%%%%%%%%%%%%%%%%%%%%%%%%%%%%%%%%%%%%%

An approximation 
for $S_1$ relying on hyperasymptotics is given by 
\cite{olde1995calculation}:
\begin{align}
\begin{split}
	\label{eq:hyper_stokes_constant}
S_1 \approx& 2 \pi \imu\,\coeff^{(0)}_{N_0}\left( 
    \sum_{m=0}^{\lfloor N_0/2 \rfloor-1} 
    \frac{\coeff^{(1)}_m \Gamma(N_0+\beta-m) }
    {A^{N_0+\beta-m}}\right)^{-1}\\ 
    \approx& %5.4703\times 10^{-3}\,\imu;
    0.0054702985252105887650131350053326816463990385103064244677326162\imu .
    \end{split}
\end{align}
We did compute $S_1$ with \eqref{eq:hyper_stokes_constant} with an
accuracy of $\bigo{10^{-65}}$ using $N_0=200$. \abbeq
\eqref{eq:hyper_stokes_constant} requires knowledge of
the coefficients of
both the perturbative and the first non-perturbative sector.
Note that it is possible to compute $S_1$ without knowing
the coefficients of the first non-perturbative sector
using the so-called large-order relations
\begin{equation}
	\label{eq:large_order_behaviour}
	\coeff^{(0)}_n \sim \frac{\Gamma(n+\beta)}{A^{n+\beta}} \, S_1 \,
	\left( \coeff^{(1)}_0 + \bigo{n^{-1}}\right), \text{as}\quad n\to \infty
\end{equation}
The leading order behaviour in \eqref{eq:large_order_behaviour} provides a
sequence which converges to $S_1$ as $\bigo{n^{-1}}$ and involves only
the free coefficient $\coeff^{(1)}_0$ from the first non-perturbative sector,
which \textit{defines} the Stokes constant. The value in \abbeq 
\eqref{eq:hyper_stokes_constant} corresponds to the choice $\coeff^{(1)}_0=3/2$,
which we did so value of the Stokes constant is the same as 
in \cite{aniceto_resurgence_2016,bacsar2015hydrodynamics,heller2015hydrodynamics}.
There is a connection between the value of the
Stokes constant and the ambiguity in the value of the
parameter $\sigma$. The positive real axis is a Stokes line,
meaning that the Borel transform
$\borel{\Phi}(\xi)$
has
branch-cut singularities at the locations
$\xi = A, \, 2A, \, 3A \dots$.
Therefore the definition 
\eqref{eq:inverse_borel_pade}
is ambiguous in the choice of angle
$\theta$.

When we move
the integration path across the
Stokes line from below 
and thus increase the angle
$\theta$ in \eqref{eq:inverse_borel_pade}
from $\theta_-=-\eps$ to 
$\theta_+=+\eps$ we get a discontinuity
in the result of the Borel resummation
\eqref{eq:inverse_borel_pade}. Crossing
the Stokes line in 
\eqref{eq:inverse_borel_pade} 
while keeping the
value of the transseries parameter
$\sigma$ from \eqref{eq:transseries_f}
constant corresponds to moving
from one Riemann sheet to the other.
Alternatively, we can alter the value
of the transseries parameter as 
$\sigma \to \sigma - S_1$ in order
to cancel the discontinuity. We require
the result of the resummation for the
whole transseries \eqref{eq:transseries_f}
to be real-valued on the positive
real axis, which is known as
Median-resummation. The reality
constraint fixes the
imaginary part of the transseries
parameter $\sigma$. Median resummation
requires\footnote{For more 
details see \textit{e.g.} 
the review \cite{Aniceto:2018bis}}
\begin{equation}
    \label{eq:im_part_sigma}
    \imu \, \mathbb{I}\text{m}(\sigma) = \pm \frac{S_1}{2}, \quad \text{for} \mp \theta >0.
\end{equation}
If we choose a convention on the path
along which we carry out the integration in
\eqref{eq:inverse_borel_pade}
(below/above the
real axis in the Borel plane), the only degree of 
freedom that is left is
the real part of the parameter $\sigma$, 
which makes sense
given that we have a one-parameter
family of real solutions.

%%%%%%%%%%%%%%%%%%%%%%%%%%%%%%%%%%%%%%%%%%%%%%%%
\section{Coefficient functions
\texorpdfstring{$F_r(W)$}{Fr(W)}}
\label{ap:calculation_f_r}
%%%%%%%%%%%%%%%%%%%%%%%%%%%%%%%%%%%%%%%%%%%%%%

The ODEs (\ref{eq:recursion_f_r}) can be rewritten as
\begin{equation}
	\label{eq:recursion_f_r_2}
	\mathcal{L} F_r(W) = g_r(W); \quad \quad \quad \mathcal{L} \equiv (1+W)W \diff{}{W}-1,
\end{equation}
where the homogeneous equation $\mathcal{L}F_r(W)=0$ is the
same for all $F_r$, and $g_r(W)$ is the inhomogeneity
which does depend on the functions $\{F_s\,|\,s\leq r-1 \}$ and their derivatives.
 It is easy to 
check that the function $\Theta(W)=W/(1+W)$ solves the homogeneous equation $\mathcal{L}\Theta = 0$.
This motivates rescaling the $F_r$ to simplify the left-hand-side of (\ref{eq:recursion_f_r_2}):
\begin{equation}
	\label{eq:rescaling_f_r}
	F_r(W) = \frac{W}{1+W} Y_r(W); \quad \quad\quad \mathcal{L}F_r(W) = W^2 Y_r^\prime(W);
\end{equation}
The advantage of working with $Y_r(W)$ is that we
can give an explicit formula for the
solutions:
\begin{equation}
\label{eq:solution_to_recursion_y_r}
Y_r(W) = \int\, \diffd W \, W^{-2} g_r(W)+c_r;
\end{equation}
The integrand of (\ref{eq:solution_to_recursion_y_r}) is
found to be given by the recurrence relation
\begin{align}
\label{eq:yr_prime_recurrence}
Y_r^\prime(W) =W^{-2} g_r(W) =& \frac{(4-\beta)(1+W)}{A W^2} \delta_{1,r}
-\frac{8}{A W} Y_{r-1}(W) \\
&+\sum_{k=0}^{r-1} \bigg[ 
\frac{9-r}{2A(1+W)} Y_k(W) Y_{r-k-1}(W)  \\
&+ \frac{1}{\left(1+W\right)^3} \, Y_k(W) 
\left( \frac{\beta}{A} Y_{r-k-1}(W) - (1-\delta_{k,0}) Y_{r-k}(W)\right)\\
&+ \frac{W}{\left(1+W\right)^2}\, Y_k(W) 
\left( \frac{\beta}{A} Y_{r-k-1}^\prime(W) - (1-\delta_{k,0}) Y_{r-k}^\prime(W)\right)
\bigg].
\end{align}
Adding an integration constant $c_r$ in (\ref{eq:solution_to_recursion_y_r}) corresponds 
to  adding a multiple of the function $\Theta(W)$,
$F_r(W) \to F_r(W) + c_r \Theta(W)$. In general, the rational
decomposition of the integrand in (\ref{eq:solution_to_recursion_y_r})
includes a term of order $W^{-1}$, which leads to logarithms in
the $Y_r(W)$. There is a unique choice 
of the set $\{c_r\,|\,r \geq0\}$ for which the $Y_r(W)$
are rational functions in $W$ without any logarithmic terms. Once the
$Y_r$ have been computed, the functions
$F_r$ are easily obtained by multiplying the $Y_r$ with
the factor $W (1+W)^{-1}$. Our method allows us to compute
as many functions $F_r$ as we want. The first few functions
are given by:
\begin{align}
\label{eq:f_r_functions_first_few}
\begin{split}
F_0(W) =& 1+W ;\qquad F_1(W)=\frac{2 W^3+(\beta +4) W^2+\beta  (\beta +7) W+\beta}{A(1+W)} ;\\
F_2(W) =&\frac1{2 A^{2}\left(1+W\right)^{3}}\Bigl( 4 W^{6}+\left(7 \beta +22\right) W^{5}+\left(8 \beta^{2}+70 \beta +32\right) W^{4}+\left(\beta^{3}+29 \beta^{2}+145 \beta +10\right) W^{3}\Bigr.\\
&\Bigl.+\left(2 \beta^{3}+34 \beta^{2}+110 \beta -8\right) W^{2}+\left(\beta^{4}+11 \beta^{3}+34 \beta^{2}+10 \beta \right) W +2
\Bigr);\\
F_3(W) =&\frac1{6 A^{3}\left(1+W\right)^{5}}\Bigl( 6 W^{9}+\left(26 \beta +60\right) W^{8}+\left(45 \beta^{2}+317 \beta +210\right) W^{7}\Bigr.\\
&+\left(30 \beta^{3}+438 \beta^{2}+1212 \beta +336\right) W^{6}+\left(11 \beta^{4}+252 \beta^{3}+1677 \beta^{2}+2050 \beta +254\right) W^{5}\\
&+\left(44 \beta^{4}+708 \beta^{3}+3210 \beta^{2}+1522 \beta +92\right) W^{4}+\left(72 \beta^{4}+903 \beta^{3}+3054 \beta^{2}+297 \beta +42\right) W^{3}\\
&+\left(-2 \beta^{6}-30 \beta^{5}-82 \beta^{4}+410 \beta^{3}+1524 \beta^{2}-220 \beta +48\right) W^{2}\\
&\Bigl.+\beta  \left(\beta^{5}+15 \beta^{4}+86 \beta^{3}+188 \beta^{2}-36 \beta +68\right) W -18 \beta^{2}+12 \beta\Bigr) ;
\end{split}
\end{align}

Note that we have not made a distinction between $F_r(\tau)$ and $F_r(W(\frac32 \tau))$
to keep our notation simple. In order to obtain the original transasymptotic
coefficient functions $F_r( \tau)$ from  (\ref{eq:transas_summation}), the variable $W$
in (\ref{eq:f_r_functions_first_few}) must be replaced by $W(\frac32\tau)$.

%%%%%%%%%%%%%%%%%%%%%%%%%%%%%%%%%%%%%%%%%%
\section{Coefficients of 
\texorpdfstring{$\gamma(w)$}{gamma(w)}}
\label{ap:gamma_perturbative}
%%%%%%%%%%%%%%%%%%%%%%%%%%%%%%%%%%%%%%%%%

We give the first four coefficients of the perturbative
expansion of $\gamma(w)$ (see \eqref{eq:gamma_zeroth_ord}),
\begin{equation}
    \label{eq:gamma_pert_expansion}
    \gamma(w) = \sum_{n=0}^{\infty} \gamma_n w^{-n}.
\end{equation}
This expansion solves
\eqref{eq:transasymptotics_is_constant}.
To simplify the notation, let us define 
\begin{equation}
    c\equiv f_\text{ac}(w_0) -1,
\end{equation}
where $f_\text{ac}(w_0)$ is the (numerical)
analytical 
continuation of $f(w)$ from $w=0$
to $w=w_0$, as 
explained in Section \ref{sec:transasymptotics}.
The first four coefficients $\gamma_n$ are then given 
by
\begin{align}
    \label{eq:gamma_first_three_coeffs}
    \begin{split}
    \gamma_0 =& \frac23 c\, e^c ;\\
    \gamma_1 =&-\frac{2e^c}{3A} \left(\beta +2 c^3+(\beta +4) c^2+\beta  (\beta +7) c\right);\\
    \gamma_2 =&\frac{e^c }{3 A^2}\bigg(2 \beta  \left(\beta ^2+7 \beta -1\right)+4 c^5+4 (\beta +7) c^4+\left(5 \beta ^2+41 \beta
   +38\right) c^3\\
   &+2 \left(\beta ^3+12 \beta ^2+34 \beta +4\right) c^2+\beta  \left(\beta ^3+15 \beta ^2+55 \beta
   +10\right) c\bigg);\\
   \gamma_3 =& -\frac{e^c}{9 A^3} \bigg(3 \beta  \left(\beta ^4+15 \beta ^3+51 \beta ^2-18 \beta +4\right)+8 c^7+12 (\beta +10) c^6 \\
   &+6\left(3 \beta ^2+33 \beta +83\right) c^5+\left(13 \beta ^3+207 \beta ^2+872 \beta +620\right) c^4\\
   &+3 \left(3
   \beta ^4+52 \beta ^3+261 \beta ^2+349 \beta +78\right) c^3\\
   &+3 \left(\beta ^5+21 \beta ^4+142 \beta ^3+321 \beta
   ^2+74 \beta +16\right) c^2\\
   &+\beta  \left(\beta ^5+24 \beta ^4+188 \beta ^3+507 \beta ^2+198 \beta +20\right)
   c\bigg).
    \end{split}
\end{align}

%%%%%%%%%%%%%%%%%%%%%%%%%%%%%%%%%%
\section{Lambert-W function}
\label{ap:lambert_w}
%%%%%%%%%%%%%%%%%%%%%%%%%%%%%%%%%%

The Lambert-W function (see \cite[\href{http://dlmf.nist.gov/4.13}{\S 4.13}]{NIST:DLMF})
is defined as the solution to the equation
\begin{equation}
\label{eq:def_lambert_w}
W(z) \eul^{W(z)} = z .
\end{equation}
The function $W(z)$ has infinitely many branches, which are
known as $W_k(z)$, where $k$ is an integer. Only two
of those branches, $W_{-1}(z)$ and $W_0(z)$, 
return real values on subsets of the real line. In the case of
our problem, the MIS equation \eqref{eq:ode_mis_model_f_of_w},
the Lambert-W function appears in the context of the
transasymptotic summation \eqref{eq:transas_summation},
where the leading-order contribution in $w^{-1}$ is given by
 \begin{equation}
 \label{eq:f0_evaluated}
 F_0(\tau(w)) = 1+W\left(\frac32 \sigma
w^\beta \eul^{-A w}\right).
\end{equation}
As $w \to +\infty$ we require $f(w) \to 1$. This means
that
 $W(\dots) \to 0$ in \eqref{eq:f0_evaluated}. For
 $k \neq0$ the branches $W_k(z)$ diverge as $z \to 0$.
Therefore, we need to choose the
branch $W_0$ at $w = +\infty$, which admits the Taylor
expansion $W_0(z) = z + \dots$ around $z=0$ and is hence
consistent with the behaviour of $f(w)$ near $w=+\infty$. 
For large
arguments, the branch $W_0$ admits the following expansion
\cite{corless_lambertw_1996}:
\begin{equation}
    \label{eq:expansion_lambert_w}
    W_0(z) = L_1 - L_2 + \sum_{k=0}^{\infty} \sum_{m=1}^{\infty} \, 
    C_{k m} L_1^{-(k+m)} \, L_2^m, \,
\end{equation}
where
\begin{align}
    \begin{split}
    L_1 &= \log w \,; \\
    L_2 &= \log (\log  w)\,; \\
    C_{k m} &= \frac{(-1)^{k+m+1}}{m!} \text{Stir}(k+m,k+1)\, .
    \end{split}
\end{align}
The expression $\text{Stir}$ denotes Stirling circle
numbers of the first kind. The presence of
logarithmic terms
in the expansion \ref{eq:expansion_lambert_w} explain
how logarithmic terms arise in the transseries $\Psi$ in
\eqref{eq:p4_family_solutions} from the transseries $\mathcal{F}$
in \eqref{eq:transseries_f} when going from $w= +\infty$ to
$w=-\infty$. Note that the magnitude of the
 exponential scale $\tau\sim \eul^{-A w}$
changes from small to large when
the sign of $w$ is flipped from $(+)$ to $(-)$, which makes it necessary
to use the expansion \eqref{eq:expansion_lambert_w}.
Let us also note that the
Lambert-W function has a square root branch point at
$z=-\eul^{-1}$.

%%%%%%%%%%%%%%%%%%%%%%%%%%%%%%%%%%%%%%%%%
\section{Taylor-series method}
\label{ap:taylor}
%%%%%%%%%%%%%%%%%%%%%%%%%%%%%%%%%%%%%%%%%

In the Taylor-series method (see \cite[\href{http://dlmf.nist.gov/3.7.ii}{\S 3.7(ii)}]{NIST:DLMF})
we combine, at a regular point $w=w_0$, the Taylor series 
$f(w)=\sum_{n=0}^\infty b_n\left(w-w_0\right)^n$ with our differential equation
\eqref{eq:ode_mis_model_f_of_w} and obtain the recurrence relation
\begin{align}
    \begin{split}\label{eq:taylor_coeff_recurrence}
    w_0(n+1)b_0b_{n+1}=&A\delta_{n,1}+(Aw_0+\beta-4)\delta_{n,0}-\tfrac12 w_0(n+1)\sum_{m=1}^n
    b_mb_{n+1-m}\\
    &-\tfrac12 (n+8)\sum_{m=0}^nb_mb_{n-m}-Ab_{n-1}-(Aw_0-8)b_n,\qquad n\geq0.
    \end{split}
\end{align}
With this method it is very easy to `walk' in the complex $w$-plane. 
Once we know $b_0=f(w_0)$
(either from a local expansion at the origin, or a branch-point, or from the asymptotic expansion)
we can compute many coefficients in the Taylor-series expansion, and use this Taylor series
to make a small step in the complex $w$-plane, that is, compute $f(w_0+h)$ and use this as
the new $b_0$.

%\bibliographystyle{plain}
%\bibliography{MIS_references.bib}

\printbibliography %Prints bibliography
\end{document}